%% file: Main.tex
\documentclass[aps,prl,twocolumn,superscriptaddress,10pt,showkeys]{revtex4-2}

\usepackage[utf8]{inputenc}
\usepackage{amsmath,amssymb,amsthm,amsfonts}
\usepackage{mathtools}
\usepackage{physics}
\usepackage{graphicx}
\usepackage{subcaption}
\usepackage{bm}         
\usepackage{placeins}

\usepackage{tikz}
\usetikzlibrary{decorations.markings}
\usepackage{xcolor}
\definecolor{my_purple}{RGB}{150, 0, 200}
\definecolor{my_orange}{HTML}{FF7F50}
\input{diagram_switch}

\input{diagram_superposition}
\input{diagram_coc}
\input{diagram_cos}
\input{diagram_soc}
\input{diagram_sos}
\input{plotlegend}

\usepackage{hyperref}
\graphicspath{{./figures/}}
\hypersetup{
    colorlinks=true,
    linkcolor=black,
    citecolor=magenta,
    urlcolor=blue,
}

\begin{document}

\title{Communication through the combination of quantum switch and coherent superposition of channels}

\author{Arghyabindu Patra}
\email{arghyabindu1@gmail.com}
\affiliation{School of Physics, University of Hyderabad, Hyderabad, India}
\affiliation{Center for Quantum Science and Technology, Siksha 'O' Anusandhan University, Bhubaneswar-751030, India}

\author{Abdul Q Batin}
\email{aqbatin@gmail.com}
\affiliation{Center for Quantum Science and Technology, Siksha 'O' Anusandhan University, Bhubaneswar-751030, India}

\author{Prasanta K. Panigrahi}
\email{pprasanta@iiserkol.ac.in}
\email{director.cqst@soa.ac.in}
\affiliation{Center for Quantum Science and Technology, Siksha 'O' Anusandhan University, Bhubaneswar-751030, India}

\begin{abstract}
The quantization of particle trajectories gives rise to remarkable features such as the coherent superposition of quantum channels and the quantum switch, which offer significant advantages in the communication of both classical and quantum information. In this study, we investigate the classical and quantum capacities of various supermaps, including individual quantum switches, coherent superpositions of channels, their combinations, and hybrid superpositions. A comparative analysis of these configurations reveals the scenarios in which specific combinations yield enhanced communication advantages.
\end{abstract}

\keywords{Quantum Switch, Coherent Superposition, Classical Capacity, Quantum Capacity}

\maketitle

\section{Introduction}
One of the major obstacles in the development of cutting-edge quantum technologies is the presence of noise in quantum channels. Since most quantum systems operate at extremely low temperatures, even minimal external radiation can disturb their fragile qubit states. In the context of quantum communication, such noise severely degrades the information transmission rate. Nevertheless, several theoretical and experimental efforts are underway to address these challenges through both algorithmic and hardware-based approaches. As achieving a perfectly noise-free quantum system is nearly impossible, techniques such as quantum error mitigation \cite{cai2023quantum}  and quantum error correction \cite{devitt2013quantum} have emerged as promising solutions to counteract the detrimental effects of noise.

Over the past two decades, a growing body of research has emerged around the idea of treating spacetime as a quantum entity, an approach that has gained strong interest in the scientific community. In particular, recent advances have shown that quantum path—the coherent superposition of quantum operations and superposition of quantum operations with alternative causal order—can give remarkable advantages. These benefits span a wide range of applications, including quantum information processing \cite{chiribella2012perfect,araujo2017purification,brukner2015bounding,kechrimparis2024enhancing,kaur2023remote}, quantum computation \cite{chiribella2013quantum,colnaghi2012quantum,araujo2014computational,araujo2017quantum,apadula2024no}, quantum metrology \cite{zhao2020quantum,chapeau2021noisy,an2024noisy,mothe2024reassessing,goldberg2023evading,yin2023experimental}, and quantum thermodynamics \cite{felce2020quantum,dieguez2023thermal,capela2023reassessing,francica2022causal,zhu2023charging,cao2022quantum,zhao2022influence}. 

In this study, we offer a detailed investigation of the impact of path superposition and the quantum switch with various configuration like quantum switch of two quantum switch, coherent superposition of two coherent superposition of channels, quantum switch of two coherent superposition of channels, coherent superposition of two quantum switch, on the maximum rate of error-free communication. This line of research gained considerable attention following the pioneering work of Chiribella et al. \cite{ebler2018enhanced}, who demonstrated that a zero-capacity depolarizing channel can effectively act as a perfect channel when embedded within a quantum switch configuration. A. A. Abbott et al. \cite{abbott2020communication} conducted a comparative analysis of these two frameworks in terms of their information transmission capacities. Recently various studies has been reported on the enhancement of quantum capacity \cite{bhaskar2020experimental,hao2021entanglement,sk2024information}

In the preliminaries section, the framework for the superposition of channels has been discussed. The following section discusses the supermaps composed of a combination of different types of quantum channel superposition. In capacity section, the theory and analysis of different capacities is established, and in the result section, the capacities of different superpositions of quantum channels are shown. In the discussion, the advantages and disadvantages of this different superposition have been studied.

\section{Preliminaries}

Since the emergence of quantum Shanon theory, the evolution of qubits or quantum information in various trajectories has received considerable attention. Unlike classical information, quantum information can evolve in multiple trajectories simultaneously. It gives rise to the second-quantized Shannon theory. This theory says that not only are information carriers quantized, but also that trajectories can be quantized. The early study of trajectories is done for isolated or closed systems. However, in a real lab experiment, these evolutions are open systems or interact with the environment. The proper framework for studying the simultaneous evolution of multiple trajectories or paths is given by \citet{chiribella2019quantum}.

\subsection{Framework}
The framework is based on the theory of the supermap, also known as higher-order transformation \cite{chiribella2008transforming}. Any quantum evolution, whether it is time evolution, interaction, or any other type, can be expressed as a quantum operation $\mathcal{E}$ \cite{NielsenChuang2010}. The evolution can also be represented as

\begin{equation}
    \rho_{\text{out}}(\mathcal{H}_{out}) = \mathcal{E}(\rho_{\text{in}}(\mathcal{H}_{in}))
    \label{eq:quantum_evolution}
\end{equation}

Here, $\rho_{\text{in}}$ is the input system, which goes under the quantum operation $\mathcal{E}$, which is a CPTP map, and after evolution, the system becomes $\rho_{\text{out}}$. The Hilbert space associated with $\rho_{\text{in}}$ is $\mathcal{H}_{in}$ $\&$ $\mathcal{H}_{out}$ is the Hilbert space associated with $\rho_{\text{out}}$. The map is entirely positive and trace preserving. A completely positive trace preserving (CPTP) map is commonly referred to as a quantum channel. Kraus operators can represent the quantum channel \cite{choi1975completely}, which is helpful in directly representing the noise in the system.

\begin{equation}
\mathcal{E}(\rho) = \sum_{i} K_i \rho K_i^{\dagger}
\label{eq:kraus_representation0}
\end{equation}

The collection of Kraus operators, $\{ K_i \}$, holds the completeness relation $\sum_{i}{ K_i^{\dagger}  K_i} = I $.

The quantum noise disturbs the qubit; disturbance means fluctuations or rotations of the qubit. These disturbances can be mathematically modeled as transformations on the Bloch sphere. If the quantum channel has noise that only rotates the qubit around the $X$-axis, then the result is known as a bit-flip channel. Similarly, rotation around the $Z$-axis defines the phase-flip channel. $Y$-axis rotation can be define with $Z$-axis $\&$ $X$-axis rotation. For the most general case, or the practical case, a quantum channel has all $X$-axis,$Y$-axis $\&$ $Z$-axis rotation; this type of channel is called a Pauli channel \cite{NielsenChuang2010}.

Standard quantum information processing can be described using the quantum circuit framework. However, in the case of indefinite causal order, where the sequencing of quantum operations remains undetermined, the standard quantum circuit framework is not helpful. To describe an indefinite causal order, which is also known as a quantum switch, we need the supermap framework. The quantum supermap considers a larger system where it maps one quantum channel to another quantum channel, so it is a map of a map \cite{chiribella2008transforming}. 

 \begin{align*}
 \text{Quantum Channel: } \quad & \mathcal{E}: \rho_{in}\to \rho_{out} \\
\text{Quantum Supermap: } \quad & \mathcal{S}:  \mathcal{E}    \mapsto \mathcal{E}' \\
\end{align*}

Here, $\mathcal{E}$ is a map that maps an initial quantum state $\rho_{in}$ to its final form $\rho_{out}$ following application of a quantum process. $\mathcal{S}$ is a supermap that acts as a higher-level operator that maps the quantum channel $\mathcal{E}$ to another quantum channel $\mathcal{E}'$.

\subsection{Quantum Switch}

The quantum switch involves two distinct channels, denoted $\boldsymbol{\mathcal{E}}^{(1)}$ $\&$ $\boldsymbol{\mathcal{E}}^{(2)}$. And, which are entangled with a control qubit $\rho_c$. The information passes through $\boldsymbol{\mathcal{E}}^{(1)}$ and then $\boldsymbol{\mathcal{E}}^{(2)}$ when the control qubit is in $\dyad{0}{0}$. Conversely, if the control qubit is in state $\dyad{1}{1}$, the information goes through $\boldsymbol{\mathcal{E}}^{(2)}$ and then $\boldsymbol{\mathcal{E}}^{(1)}$. But when the control is in a state of $\dyad{0}{1}$ or $\dyad{1}{0}$, the information passes in both paths simultaneously, which creates a superposition of alternative causal order. In the Fig.~\ref{fig:allquantumswitch}, the schematic representation of a quantum switch is shown as a diagram.  

\begin{figure}[htbp]
    \centering

    \begin{subfigure}[b]{0.48\columnwidth}
        \centering
        \textbf{(a)}
        \vspace{2mm}
        \diagramswitchaa
        \label{fig:quantum_switch_a}
    \end{subfigure}
    \hfill % This command pushes the two subfigures apart
    \begin{subfigure}[b]{0.48\columnwidth}
        \centering
        \textbf{(b)}
        \vspace{2mm}
        \diagramswitchbb
        \label{fig:quantum_switch_b}
    \end{subfigure}

    \vspace{5mm}

    \begin{subfigure}[b]{\columnwidth}
        \centering
        \textbf{(c)}
        \vspace{2mm}
        \diagramswitch
        \label{fig:quantum_switch}
    \end{subfigure}

    \caption{This is a diagram of a quantum switch. (a) In the case where the control is in the $\dyad{0}{0}$ state, the information passes $\boldsymbol{\mathcal{E}}^{(1)}$ before proceeding to $\boldsymbol{\mathcal{E}}^{(2)}$. (b) In this case, the control is in $\dyad{1}{1}$ state, the order is $\boldsymbol{\mathcal{E}}^{(2)}$ then $\boldsymbol{\mathcal{E}}^{(1)}$. (c) For control qubits in the superposition $\ketbra{+}{+}$, the information passes concurrently through a superposition of the two pathways at the same time.  $\boldsymbol{\mathcal{E}}^{(1)}$ $\&$ $\boldsymbol{\mathcal{E}}^{(2)}$ are two quantum channels.}
    \label{fig:allquantumswitch}
\end{figure}

The quantum supermap comes in handy to represent the quantum switch operation. With the help of a quantum supermap, we can construct a quantum channel out of quantum channel $\boldsymbol{\mathcal{E}^{(1)}}$ $\&$ $\boldsymbol{\mathcal{E}^{(2)}}$ which can mimic overall quantum switch operation with just a single quantum channel \cite{chiribella2013quantum}.
The Kraus operator of quantum channel $\boldsymbol{\mathcal{E}^{(1)}}$ can be written as  $\mathcal{E}^{(1)}(\rho) = \sum_{i} K_i \rho K_i^{\dagger}$ and for the quantum channel $\boldsymbol{\mathcal{E}^{(1)}}$, $\mathcal{E}^{(2)}(\rho) = \sum_{j} L_j \rho L_j^{\dagger}$. Here $\{ K_i \}$ $\&$ $\{ L_j \}$  are the set of Kraus operator, satisfy $\sum_{i}{ K_i^{\dagger}  K_i} = I $, $\&$ $\sum_{j}{ L_j^{\dagger}  L_j} = I$.
The overall channel Kraus operator of the quantum switch can be constructed with the Kraus operator of quantum channel $\boldsymbol{\mathcal{E}^{(1)}}$ $\&$ $\boldsymbol{\mathcal{E}^{(2)}}$ \cite{ebler2018enhanced}.
\begin{equation}
M_{ij} = L_j K_i \otimes \dyad{0}{0} + K_i L_j \otimes \dyad{1}{1}  
\label{eq:kraus_representation3}
\end{equation}

The overall quantum switch can now be described with a single channel $\boldsymbol{\mathcal{S}}$, having Kraus operators $M_{ij}$.

\begin{equation}
\mathcal{S} \left(\mathcal{E}^{(1)},\mathcal{E}^{(2)} \right) \left(\rho_t \otimes \rho_c \right) = \sum_{i,j} M_{ij} \left(\rho_t \otimes \rho_c \right) M_{ij}^{\dagger}
\label{eq:overall_switch}
\end{equation}

\subsection{Coherent Superposition of Quantum Channel}

\begin{figure}[htbp]
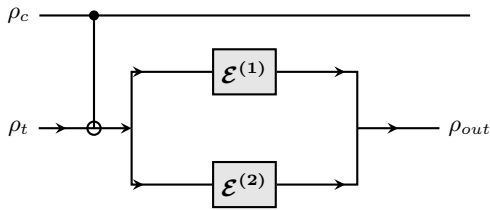

    \centering
    \diagramcoherent
    \caption{A diagram of coherent Superposition of two individual quantum channels defined as $\boldsymbol{\mathcal{E}^{(1)}}$ and $\boldsymbol{\mathcal{E}^{(2)}}$. $\rho_c$ is the control qubit, which is initiated in the state $\dyad{+}{+}$, and $\rho_t$ is the target qubit, on which the information that needs to be transferred is encoded. }
    \label{fig:Coh_supeposition}
\end{figure}

Fig.~\ref{fig:Coh_supeposition} shows a schematic diagram of the coherent superposition of two channels. Similarly like quantum switch, for control is in the state $\dyad{0}{0}$ the information passes through the channel $\boldsymbol{\mathcal{E}^{(1)}}$ and when control is $\dyad{1}{1}$ in the state the information passes through $\boldsymbol{\mathcal{E}^{(2)}}$ only. So, constructing a supermap with this logic is problematic because we miss the other channel in each case. So, to create a larger space where two channels are always acting, we can introduce a vacuum notion. Conversely, if the control is in the  state $\dyad{0}{0}$ the information passes through the channel $\boldsymbol{\mathcal{E}^{(1)}}$ and the vacuum passes through $\boldsymbol{\mathcal{E}^{(2)}}$ and, when the control is in the state $\dyad{1}{1}$, vice versa \cite{chiribella2019quantum}.

In the quantum switch configuration, regardless of the control state, both quantum channels are always active, and therefore, the vacuum notion is unnecessary for the quantum switch. 

The paper \citet{chiribella2019quantum} defines how to construct a vacuum extension of a quantum channel. A quantum channel $\boldsymbol{\mathcal{E}^{(1)}}$ has Kraus operators $K_i$, then the Kraus operator of its vacuum extension quantum channels is

\begin{equation}
\tilde{K}_i = K_i \oplus \gamma_i \dyad{Vac}{Vac}  
\label{eq:vacuum_kraus}
\end{equation}

Here, $\tilde{K}_i$ is the Kraus operator of an extended quantum channel. $\gamma_i$ are the vacuum amplitude of $\tilde{K}_i$. Which satisfies the condition, $\sum_i |\gamma_i|^2 = 1$. To find the vacuum amplitude for a quantum channel is an engineering problem. The user must choose the vacuum amplitude, which must satisfy the given normalization condition.

The vacuum extended quantum channel of $\boldsymbol{\mathcal{E}^{(1)}}$ is $\boldsymbol{\mathcal{\tilde{E}}^{(1)}}$, Which has Kraus operator $\tilde{K}_i$  $\&$ the vacuum amplitude $\alpha_i$ which satisfies the condition $\sum_i |\alpha_i|^2 = 1$. And they are connected as
$
\tilde{K}_i = K_i \oplus \alpha_i \dyad{Vac}{Vac}  
$
. Similarly, The Vacuum extended quantum channel of $\boldsymbol{\mathcal{E}^{(2)}}$ is $\boldsymbol{\mathcal{\tilde{E}}^{(2)}}$. And the Kraus operator and vacuum amplitude are $\tilde{L}_j$ $\&$ $\beta_j$. $\beta_j$ also satisfies $\sum_j |\beta_j|^2 = 1$. And they are connected as 
$
\tilde{L}_j = L_j \oplus \beta_j \dyad{Vac}{Vac}  
$.

According to \citet{chiribella2019quantum}, the Kraus operator of the quantum channel, made up of a coherent superposition of two channel $\boldsymbol{\mathcal{\tilde{E}}^{(1)}}$ $\&$ $\boldsymbol{\mathcal{\tilde{E}}^{(2)}}$ is

\begin{equation}
N_{ij} = K_i \beta_j \oplus \alpha_i L_j  
\label{eq:coh_kraus}
\end{equation}

So, the coherent superposition of two quantum channels, $\boldsymbol{\mathcal{\tilde{S}}}$, which has the vacuum extended quantum channels $\boldsymbol{\mathcal{\tilde{E}}^{(1)}}$ $\&$ $\boldsymbol{\mathcal{\tilde{E}}^{(2)}}$ can be written as

\begin{equation}
\mathcal{\tilde{S}} \left(\mathcal{\tilde{E}}^{(1)},\mathcal{\tilde{E}}^{(2)} \right) \left(\rho_t \otimes \rho_c \right) = \sum_{i,j} N_{ij} \left(\rho_t \otimes \rho_c \right) N_{ij}^{\dagger}
\label{eq:overall_coherent_sup}
\end{equation}

\subsection{Kraus Operator of Different Channel}
We are taking the bit-flip Channel, the phase-flip Channel $\&$ depolarizing channel for this investigation.

A bit-flip channel's Kraus operators can be written as 
 \begin{align*}
 K_0 = \sqrt{1-p} \quad \mathbb{I} \qquad
 K_1 = \sqrt{p} \quad \sigma_x
\end{align*}
Here $p$ is the probability that a bit-flip noise will occur, which is also called the noise parameter. So, here $p$ is the probability of Bit-Flip error or rotation around the $X$-axis error. And with $1-p$, a qubit will pass without disturbance. 

And the bit-flip channel can be written in the operator-sum representation eq.~\ref{eq:kraus_representation0}

\begin{align*}
 {\mathcal{E}}_{Bit(1)}(\rho) &= \sum_{i=0}^{1} K_i \rho K_i^{\dagger} \\
&= K_0 \rho {K_0}^{\dagger} + K_1 \rho {K_1}^{\dagger} \\
&= (1-p) \mathbb{I} \rho \mathbb{I} + (p) \sigma_x \rho \sigma_x \\
&= (1-p)  \rho  + (p) \sigma_x \rho \sigma_x \\
 \label{eq:bit_flip01}
\end{align*}

Similarly, Let's define another bit-flip channel ${\mathcal{E}}_{Bit(2)}(\rho) = \sum_{i=0}^{1} L_i \rho L_i^{\dagger}$ . The Kraus operator of quantum switch made up ${\mathcal{E}}_{Bit(1)}(\rho)$ and ${\mathcal{E}}_{Bit(1)}(\rho)$, would be according the eq.~\ref{eq:kraus_representation3}

\begin{align*}
 M_{00} &= L_0 K_0 \otimes \dyad{0}{0} + K_0 L_0 \otimes \dyad{1}{1}   \\
M_{01} &= L_1 K_0 \otimes \dyad{0}{0} + K_0 L_1 \otimes \dyad{1}{1}   \\
M_{10} &= L_0 K_1 \otimes \dyad{0}{0} + K_1 L_0 \otimes \dyad{1}{1}  \\
M_{11} &= L_1 K_1 \otimes \dyad{0}{0} + K_1 L_1 \otimes \dyad{1}{1}   \\
\end{align*}

The Bit-Flip channel characterized by 2 Kraus operators, so the overall quantum switch described by 4 Kraus operators.

To construct coherent superposition of ${\mathcal{E}}_{Bit(1)}(\rho)$ and ${\mathcal{E}}_{Bit(2)}$, vacuum amplitude of these two channels needs to find. If the vaccum amplitude of a Bit-Flip channel is $\alpha_0 = \dfrac{1}{\sqrt{2}}$ $\&$ $\alpha_1 = \dfrac{1}{\sqrt{2}}$ \cite{chiribella2019quantum}, then the Kraus operator of coherent superposition of ${\mathcal{E}}_{Bit(1)}(\rho)$ $\&$ ${\mathcal{E}}_{Bit(2)}$, according to the Eq.~\ref{eq:coh_kraus}, would be
\begin{align*}
 N_{00} &= K_0 \beta_0 \oplus \alpha_0 L_0 =  \dfrac{1}{\sqrt{2}} \left( K_0 \oplus L_0 \right)   \\
N_{01} &= K_0 \beta_1 \oplus \alpha_0 L_1  =  \dfrac{1}{\sqrt{2}} \left( K_0 \oplus L_1 \right)   \\
N_{10} &= K_1 \beta_0 \oplus \alpha_1 L_0 =  \dfrac{1}{\sqrt{2}} \left( K_1 \oplus L_0 \right)   \\
N_{11} &= K_1 \beta_1 \oplus \alpha_1 L_1  =  \dfrac{1}{\sqrt{2}} \left( K_1 \oplus L_1 \right)  \\
\end{align*}
Similarly, the Kraus operators for a quantum switch or a coherent superposition involving two independent channels—such as two phase-flip channels or a mix of bit-flip and phase-flip channels—can be derived in an analogous manner. The overall number of Kraus operators remains four in every instance, both for the quantum switch and for the coherent superposition of two channels.

The Kraus operators of the Phase-Flip channel
 \begin{align*}
 K_0 = \sqrt{1-p} \quad \mathbb{I} \qquad
 K_1 = \sqrt{p} \quad \sigma_y
\end{align*}
Here, with probability $p$, the qubit experiences a Phase-Flip error. 

But the physical quantum channels are the Pauli channels, whose Kraus operators are defined as  

And for Pauli Channel
 \begin{align*}
 K_0 &= \sqrt{1-p} \quad \mathbb{I} \qquad
 K_1 = \sqrt{p_x} \quad \sigma_x  \\
 K_2 &= \sqrt{p_y} \quad \sigma_x \qquad
 K_3 = \sqrt{p_z} \quad \sigma_x\\
\end{align*}

Here, $p$ is equal to $p_x+p_y+p_z$. To simplify the most general error, we assume all three kinds of errors have the same probability, so that the Kraus operator will be

 \begin{align*}
 K_0 = \sqrt{1-p} \quad \mathbb{I} \qquad
 K_1 = \sqrt{\dfrac{p}{3}} \quad \sigma_x\\
 K_2 = \sqrt{\dfrac{p}{3}} \quad \sigma_x \qquad
 K_3 = \sqrt{\dfrac{p}{3}} \quad \sigma_x\\
\end{align*}
This kind of Pauli channel is called a depolarizing channel. The vacuum amplitude of the depolarizing channel can be calculated for the required system by doing quantum process tomography (QPT) \cite{chuang1997prescription, o2004quantum, PhysRevLett.91.120402}.

\section{Different Supermaps}
In this section, we discuss various supermaps, focusing on their structural composition and the representation of their overall Kraus operators in terms of the Kraus operators of individual quantum channels. In the case of a quantum switch of a switch \cite{das2022quantum}, the corresponding Kraus operators and channel capacities have been previously analyzed. Here, we explore the possible combinations of two fundamental superposition frameworks namely, the quantum switch and the coherent superposition of channels. Four distinct configurations can be constructed from these two methods, which are discussed in detail below.

\subsection{Quantum Switch of Quantum Switch}
Consider two quantum switches $\boldsymbol{\mathcal{S}_1}$ $\&$ $\boldsymbol{\mathcal{S}_2}$. $\boldsymbol{\mathcal{S}_1}$ consists of two individual quantum channels  $\boldsymbol{\mathcal{E}^{(1)}}$ $\&$ $\boldsymbol{\mathcal{E}^{(2)}}$. Similarly, $\boldsymbol{\mathcal{S}_2}$ consists of two individual quantum channels  $\boldsymbol{\mathcal{E}^{(3)}}$ $\&$ $\boldsymbol{\mathcal{E}^{(4)}}$. Both have the same control qubit $\rho_c^{\prime}$ that is initialized at $\ket{+}$.

$\boldsymbol{\mathcal{S}_1}$ $\&$ $\boldsymbol{\mathcal{S}_2}$, can be written as
\begin{equation}
\mathcal{S}_1  \left(\rho_t \otimes \rho_c^{\prime} \right) = \sum_{i,j} M_{ij}^{(1)} \left(\rho_t \otimes \rho_c^{\prime} \right) {M_{ij}^{(1)}}^{\dagger}
\label{eq:quantum_switch1}
\end{equation}

\begin{equation}
\mathcal{S}_2  \left(\rho_t \otimes \rho_c^{\prime} \right) = \sum_{k,l} M_{kl}^{(2)} \left(\rho_t \otimes \rho_c^{\prime} \right) {M_{kl}^{(2)}}^{\dagger}
\label{eq:quantum_switch2}
\end{equation}

Here, $M_{ij}^{(1)}$ $\&$ $M_{ij}^{(2)}$, which are the Kraus operator of quantum switch $\boldsymbol{\mathcal{S}_1}$ $\&$ $\boldsymbol{\mathcal{S}_2}$ is defined as 

\begin{equation}
M_{ij}^{(1)} = L_j^{(2)} K_i^{(1)} \otimes \dyad{0}{0} + K_i^{(1)} L_j^{(2)} \otimes \dyad{1}{1}  
\label{eq:kraus_qs01}
\end{equation}

\begin{equation}
M_{kl}^{(2)} = L_l^{(4)} K_k^{(3)} \otimes \dyad{0}{0} + K_k^{(3)} L_l^{(4)} \otimes \dyad{1}{1}  
\label{eq:kraus_qs02}
\end{equation}

Here, $K_i^{(1)}$, $L_j^{(2)}$, $K_k^{(3)}$ $\&$ $L_l^{(4)}$, are the Kraus operator of $\boldsymbol{\mathcal{E}^{(1)}}$, $\boldsymbol{\mathcal{E}^{(2)}}$, $\boldsymbol{\mathcal{E}^{(3)}}$ $\&$ $\boldsymbol{\mathcal{E}^{(4)}}$, respectably.

\begin{figure}[htbp]
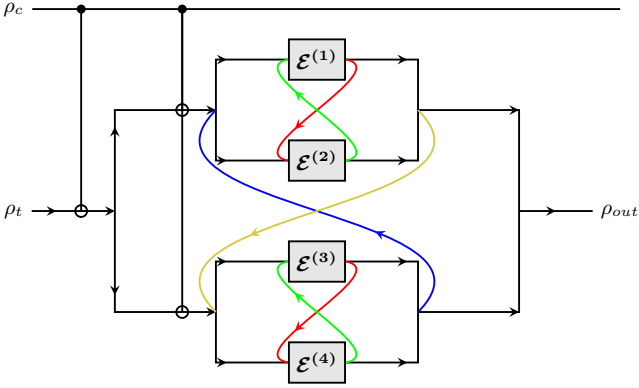

    \centering
    \diagramsos
    \caption{Diagram of a quantum switch $\mathcal{S}^{\prime}$ Consist of quantum switch $\mathcal{S}_1$ (above) and, $\mathcal{S}_2$ (below). $\rho_c$ is the control which is equal to $\dyad{++}{++}$ . $\rho_t$ is the target qubit. $\mathcal{S}_1$ has $\boldsymbol{\mathcal{E}^{(1)}}$, $\boldsymbol{\mathcal{E}^{(2)}}$ $\&$ $\mathcal{S}_2$ has $\boldsymbol{\mathcal{E}^{(3)}}$, $\boldsymbol{\mathcal{E}^{(4)}}$.}
    \label{fig:diagramsos}
\end{figure}

To construct a quantum switch made up of two quantum switches, $\boldsymbol{\mathcal{S}_1}$ $\&$ $\boldsymbol{\mathcal{S}_2}$, as shown in the Fig.~\ref{fig:diagramsos}, can be defined as 

\begin{equation}
\mathcal{S}^{\prime}  \left( \mathcal{S}_1, \mathcal{S}_2, \rho_c^{\prime \prime} \right) = \sum_{i,j,k,l} M_{ijkl}^{\prime} \left(\rho_t \otimes \rho_c \right) M_{ijkl}^{\prime \dagger}
\label{eq:quantum_sos}
\end{equation}

Here $\rho_c^{\prime \prime}$ is the control qubit of the bigger quantum switch, which is also initialized at $\ket{+}$. The inner quantum switch ($\mathcal{S}_1, \mathcal{S}_2$) has one control qubit $\rho_c^{ \prime}$ and, outer quantum switch ($\mathcal{S}^{\prime}$) has another control qubit $\rho_c^{\prime \prime}$, Overall there is two control qubit $\rho_c = \rho_c^{\prime} \otimes  \rho_c^{\prime \prime}$, which is in the state, $\dyad{++}{++}$.

$M_{ijkl}^{\prime}$ is the Kraus operator of the overall channel, which is defined as \cite{das2022quantum}

\begin{equation}
M_{ijkl}^{\prime} = M_{kl}^{(2)} M_{ij}^{(1)} \otimes \dyad{0}{0} + M_{ij}^{(1)}  M_{kl}^{(2)} \otimes \dyad{1}{1} 
\label{eq:quantum_sos_kraus}
\end{equation}

\subsection{Coherent Superposition of Coherent Superposition}
Consider two coherent superpositions, $\mathcal{\tilde{S}}_1$  $\&$ $\mathcal{\tilde{S}}_2$, Which is defined as

\begin{equation}
\mathcal{\tilde{S}}_1 \left(\rho_t \otimes \rho_c^{\prime} \right) = \sum_{i,j} N_{ij}^{(1)} \left(\rho_t \otimes \rho_c^{\prime} \right) {N_{ij}^{(1)}}^{\dagger}
\label{eq:quantum_coc01}
\end{equation}

\begin{equation}
\mathcal{\tilde{S}}_2 \left(\rho_t \otimes \rho_c^{\prime} \right) = \sum_{k,l} N_{k,l}^{(2)} \left(\rho_t \otimes \rho_c^{\prime} \right) {N_{kl}^{(2)}}^{\dagger}
\label{eq:quantum_coc02}
\end{equation}

Here $N_{ij}^{(1)}$ $\&$ $N_{kl}^{(2)}$ are the Kraus operators of coherent superposition ${\tilde{S}}_1$ $\&$  ${\tilde{S}}_2$, respectably. And the control qubit for both the coherent superposition is $\rho_c^{\prime}$, which is $ \dyad{+}{+} $. These Kraus operators are defined as

\begin{equation}
N_{ij}^{(1)} = K_i \beta_j \oplus \alpha_i L_j  
\label{eq:quantum_coc03}
\end{equation}

\begin{equation}
N_{kl}^{(2)} = K_k \beta_l \oplus \alpha_k L_l  
\label{eq:quantum_coc04}
\end{equation}

\begin{figure}[htbp]
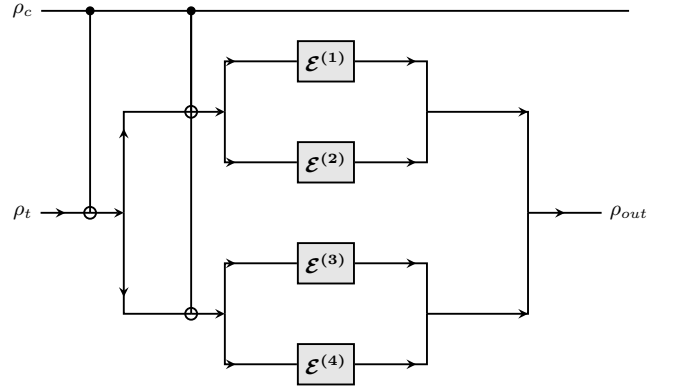

    \centering
    \diagramcoc
    \caption{A diagram of coherent superposition of coherent superposition}
    \label{fig:diagramcoc}
\end{figure}

Here, $K_i$, $L_j$, $K_k$ $\&$ $L_l$ are the Kraus opertors and, $\alpha_i$, $\beta_j$, $\alpha_k$ $\&$ $\beta_l$ are the vacuum amplitudes of channel $\boldsymbol{\mathcal{E}^{(1)}}$, $\boldsymbol{\mathcal{E}^{(2)}}$, $\boldsymbol{\mathcal{E}^{(3)}}$ $\&$ $\boldsymbol{\mathcal{E}^{(4)}}$, respectably. As shown in the Fig.~\ref{fig:diagramcoc}. The way the overall quantum switch of switch was represented, in the same coherent superposition of coherent would be, similarly, for the bigger coherent superposition $\mathcal{\tilde{S}}^{\prime}$, there is a control qubit, ${\rho_c}^{ \prime \prime}$. So the overall control qubit is a 2-qubit state, which is $\rho_c = \rho_c^{\prime} \otimes {\rho_c}^{ \prime \prime} = \dyad{++}{++}$

\begin{equation}
\mathcal{\tilde{S}}^{\prime}  \left( \mathcal{\tilde{S}}_1, \mathcal{\tilde{S}}_2, \rho_c^{\prime \prime} \right) = \sum_{i,j,k,l} N_{ijkl}^{\prime} \left(\rho_t \otimes \rho_c \right) N_{ijkl}^{\prime \dagger}
\label{eq:quantum_coc05}
\end{equation}

$N_{ijkl}^{\prime}$ is the Kraus operator of the  overall channel, which is defined as

\begin{equation}
N_{ijkl}^{(\prime)} = N_{ij}^{(1)} \beta_{kl} \oplus \alpha_{ij} N_{kl}^{(2)}  
\label{eq:quantum_coc06}
\end{equation}

$\alpha_{ij}$ and $\beta_{kl}$ are the Vacuum amplitude of these $\mathcal{\tilde{S}}_2$, $\mathcal{\tilde{S}}_2$ two channel. And it needs to be calculated.

\subsection{Quantum Switch of Two Coherent Superposition}

\begin{figure}[htbp]
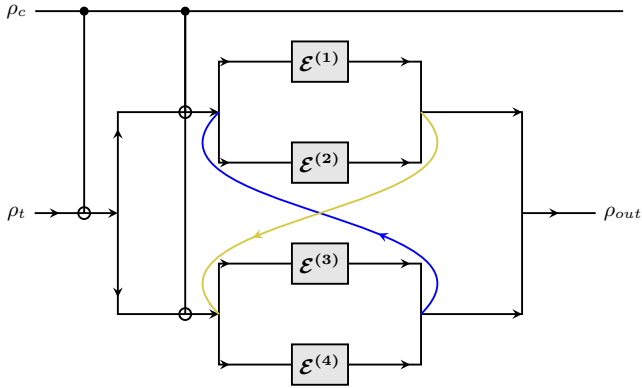

    \centering
    \diagramsoc
    \caption{A diagram of a quantum switch of coherent superposition}
    \label{fig:diagramsoc}
\end{figure}

 $\mathcal{\tilde{S}}_1$ and $\mathcal{\tilde{S}}_2$ are two coherent superposition of two quantum channels. Which can be expressed as Eq.~\ref{eq:quantum_coc01}, and Eq.~\ref{eq:quantum_coc02}. The quantum switch with these two channels, as shown in Fig.~\ref{fig:diagramsoc}, is

\begin{equation}
\mathcal{S}^{\prime}  \left(  \mathcal{\tilde{S}}_1, \mathcal{\tilde{S}}_2, \rho_c^{\prime \prime} \right) = \sum_{i,j,k,l} M_{ijkl}^{\prime} \left(\rho_t \otimes \rho_c \right) M_{ijkl}^{\prime \dagger}
\label{eq:quantum_soc}
\end{equation}

$M_{ijkl}^{\prime}$ is the Kraus opertor of overal switch, defined as

\begin{align}
    M_{ijkl}^{\prime} &= N_{kl}^{(2)} N_{ij}^{(1)} \otimes \dyad{0}{0} + N_{ij}^{(1)} N_{kl}^{(2)} \otimes \dyad{1}{1} \nonumber \\
    &= (K_k \beta_l \oplus \alpha_k L_l) (K_i \beta_j \oplus \alpha_i L_j) \otimes \dyad{0}{0} \nonumber \\
    &\quad + (K_i \beta_j \oplus \alpha_i L_j) (K_k \beta_l \oplus \alpha_k L_l) \otimes \dyad{1}{1}
\label{eq:quantum_soc_kraus}
\end{align}

\subsection{Coherent Superposition of two Quantum Switch}

\begin{figure}[htbp]
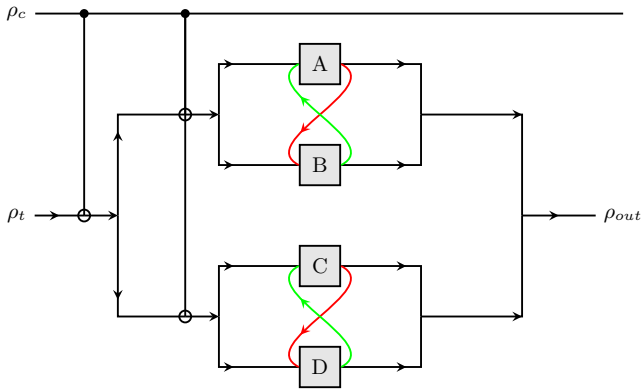

    \centering
    \diagramcos
    \caption{A diagram of coherent superposition of a quantum switch}
    \label{fig:diagramcos}
\end{figure}

$\boldsymbol{\mathcal{S}_1}$ and $\boldsymbol{\mathcal{S}_2}$ are two quantum switch, which can be expressed as Eq.~\ref{eq:quantum_switch1}, and Eq.~\ref{eq:quantum_switch2}. The coherent superposition of$\boldsymbol{\mathcal{S}_1}$ and $\boldsymbol{\mathcal{S}_2}$ as shown in Fig.~\ref{fig:diagramcos} is described as

\begin{equation}
\mathcal{\tilde{S}}^{\prime}  \left(\boldsymbol{\mathcal{S}_1}, \boldsymbol{\mathcal{S}_2}, \rho_c^{\prime \prime} \right) = \sum_{i,j,k,l} N_{ijkl}^{\prime} \left(\rho_t \otimes \rho_c \right) N_{ijkl}^{\prime \dagger}
\label{eq:quantum_cos01}
\end{equation}

$N_{ijkl}^{\prime}$ is the Kraus operator of the overall coherent superposition, which is

\begin{equation}
N_{ijkl}^{(\prime)} = M_{ij}^{(1)} \beta_{kl} \oplus \alpha_{ij} M_{kl}^{(2)}  
\label{eq:quantum_cos02}
\end{equation}

$M_{ij}^{(1)}$ $\&$ $M_{ij}^{(2)}$ are the Kraus opertor and, $\alpha_{ij}$ $\&$ $\beta_{kl}$ are the vacuum amplitude of quantum switch $\boldsymbol{\mathcal{S}_1}$  $\&$ $\boldsymbol{\mathcal{S}_2}$, respectably.

\FloatBarrier

\section{Capacity}\label{cap}

The capacity has to be calculated to check how efficient these different quantum supermaps are. It will give the threshold beyond which information can be transmitted reliably, or more precisely, how efficiently a single information transmission will happen each time.  

\subsection{Classical Capacity}

Suppose we wish to perform classical communication over a given set of quantum channels and quantum supermaps and determine the theoretical maximum transmittable information or the theoretical upper bound. In that case, it is necessary to evaluate the classical capacity of the channel \cite{schumacher1997sending, hausladen1996classical}.The present study limits its scope to the one-shot classical capacity, which quantifies the maximum amount of classical information or bits that can be transmitted through a quantum channel in a single application. The formal definition of the single-shot classical capacity is given in \cite{holevo2002capacity}

\begin{equation}
    C(\mathcal{E}) = \max_{\{p_i, \rho_i\}} \chi(\{\mathcal{E}(\rho_i), p_i\})
    \label{eq:classical_capacity}
\end{equation}

Here $C$ is the classical capacity of the quantum channel $\mathcal{E}$. $\chi$ is Holevo information, and the classical capacity is the maximum of Holevo information over a set of pure states or an ensemble of $\rho_i$ with corresponding probability $p_i$. And, the Holevo information $\chi$ is defined as

\begin{equation}
    \chi(\{\mathcal{E}(\rho_i), p_i\}) = \mathrm{S}\left(\sum_i p_i \rho_i\right) - \sum_i p_i\mathrm{S} \left( \rho_i \right)
    \label{eq:Holevo_information}
\end{equation}

Here, the first term $\mathrm{S}\left(\sum_i p_i \rho_i\right)$ is the entropy of the average state. and $\mathrm{S} \left( \rho_i \right)$ are the individual entropy. The entropy of a density matrix $\rho$ can be found as $ -\mathrm{Tr}(\rho \log_2 \rho)$ or $- \sum_i \lambda_i \log_2 \lambda_i$, here $\lambda_i$ are the eigen values of the density matrix $\rho$ \cite{vonNeumann1955}.

\subsection{Quantum Capacity}
To transmit quantum information or qubits through a quantum channel, the theoretical ultimate limit, also known as the upper bound on the amount of information that can be reliably communicated, is determined by the channel's quantum capacity \cite{lloyd1997capacity}. The single-shot quantum capacity is defined as follows:

\begin{equation}
    Q(\mathcal{E}) = \max_{\rho} I_{c}(\rho, \mathcal{E})
    \label{eq:quantum_capacity}
\end{equation}

$Q$ denotes the one-shot quantum capacity of the quantum channel $ \mathcal{E}$. $I_{c}$ is coherent information, and the quantum capacity is the maximum of coherent information over all possible inputs $\rho$, which is an optimization problem. The coherent information  $I_{c}$ defined as \cite{barnum1998information}

\begin{equation}
     I_{c}(\rho, \mathcal{E}) = \mathrm{S}\left(\mathcal{E} (\rho) \right) - \mathrm{S}_e \left( \rho , \mathcal{E} \right)
    \label{eq:coherent_information}
\end{equation}

The leading component of the coherent information is the von Neumann entropy of the quantum channel's output $\mathcal{E}$, when the input is $\ rho$. The second term signifies the entropy of the environment, also known as the complementary channel $\rho^c$, also referred to as the exchange entropy. Exchange entropy is the von Neumann entropy of the complementary channel $\rho^c$. The Kraus operator of complementary channel $\{ R_k \}$, can be constructed with the Kraus operator of the $\rho$, which is   $\{ K_i \}$, by the Stinespring dilation theorem \cite{roofeh2024noisy}.

\section{Results}
In this section, our analysis's primary findings are the classical and quantum capacities, which were mentioned in the previous section. So, basically, we have a six-channel configuration. The first is a quantum switch, whose capacities are shown in the green lines. The second configuration corresponds to the coherent superposition of two channels, depicted by the purple lines in the figure. The quantum switch of quantum switch (SoS) is shown in the black lines, the quantum switch of coherent superposition (SoC) is mentioned in the red lines, the cohesive superposition of quantum switch (CoS) is shown in the blue lines, and the coherent superposition of coherent superposition is shown as orange lines. We calculate the one-shot classical and quantum capacities for these different configurations, which are given in the appendix section.

\begin{figure}[htbp]
    \centering
    \includegraphics[width=\columnwidth]{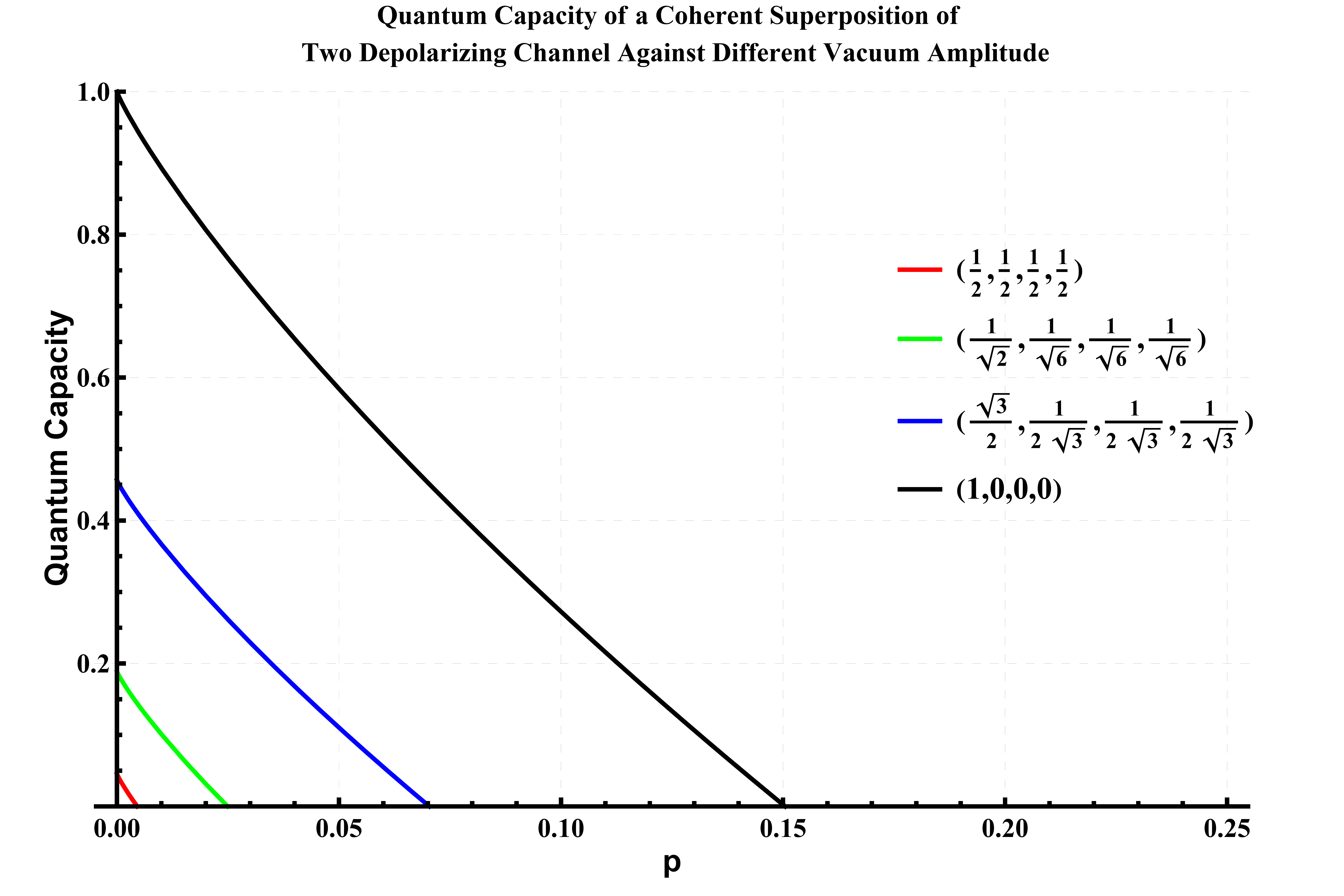}
    \caption{Quantum capacity of a coherent superposition of two depolarizing channels as a function of the depolarizing noise parameter $p$. The curves correspond to four different choices of vacuum amplitude $(\alpha_0, \alpha_1, \alpha_2, \alpha_3)$: $(\frac{1}{2}, \frac{1}{2}, \frac{1}{2}, \frac{1}{2})$ (red), $(\frac{1}{\sqrt{2}}, \frac{1}{\sqrt{6}}, \frac{1}{\sqrt{6}}, \frac{1}{\sqrt{6}})$ (green), $(\frac{\sqrt{3}}{2}, \frac{1}{2\sqrt{3}}, \frac{1}{2\sqrt{3}}, \frac{1}{2\sqrt{3}})$ (blue), and $(1, 0, 0, 0)$ (black).}
    \label{fig:vacuum_amplitude}
\end{figure}

We also examine the coherent superposition of two channels $\boldsymbol{\mathcal{E}^{(1)}}$ $\&$ $\boldsymbol{\mathcal{E}^{(2)}}$, which has same vacuum amplitude. The choice of vacuum amplitude depends upon the user. Because it is an engineering problem \cite{chiribella2019quantum}. However, the quantum capacity depends upon the choice of vacuum amplitude. In the Fig.~\ref{fig:vacuum_amplitude}, an example is given. This figure illustrates the quantum capacity associated with a coherent superposition of two depolarizing channels, evaluated across a range of vacuum amplitudes that satisfy the equation~\ref{eq:vacuum_kraus}. $(\alpha_0, \alpha_1, \alpha_2, \alpha_3)$ combines vacuum amplitude for a coherent superposition of two individual depolarizing channels.$\alpha_0$, corresponds to the vacuum amplitude of  Kraus operator, $ K_0 = \sqrt{1-p} \quad \mathbb{I}$, which define the no error operation. From the Fig.~\ref{fig:vacuum_amplitude}, it is clear that, as we increase the $\alpha_0$ value, the quantum capacity of the channel increases, so for the best result for our discussion, we took the vacuum amplitude as $(1, 0, 0, 0)$.

\subsection{Classical Capacities of Different Channels}

\begin{figure*}[tp]
    \centering
    \subfloat[\label{fig:classical01}]{\fbox{\includegraphics[width=0.32\textwidth]{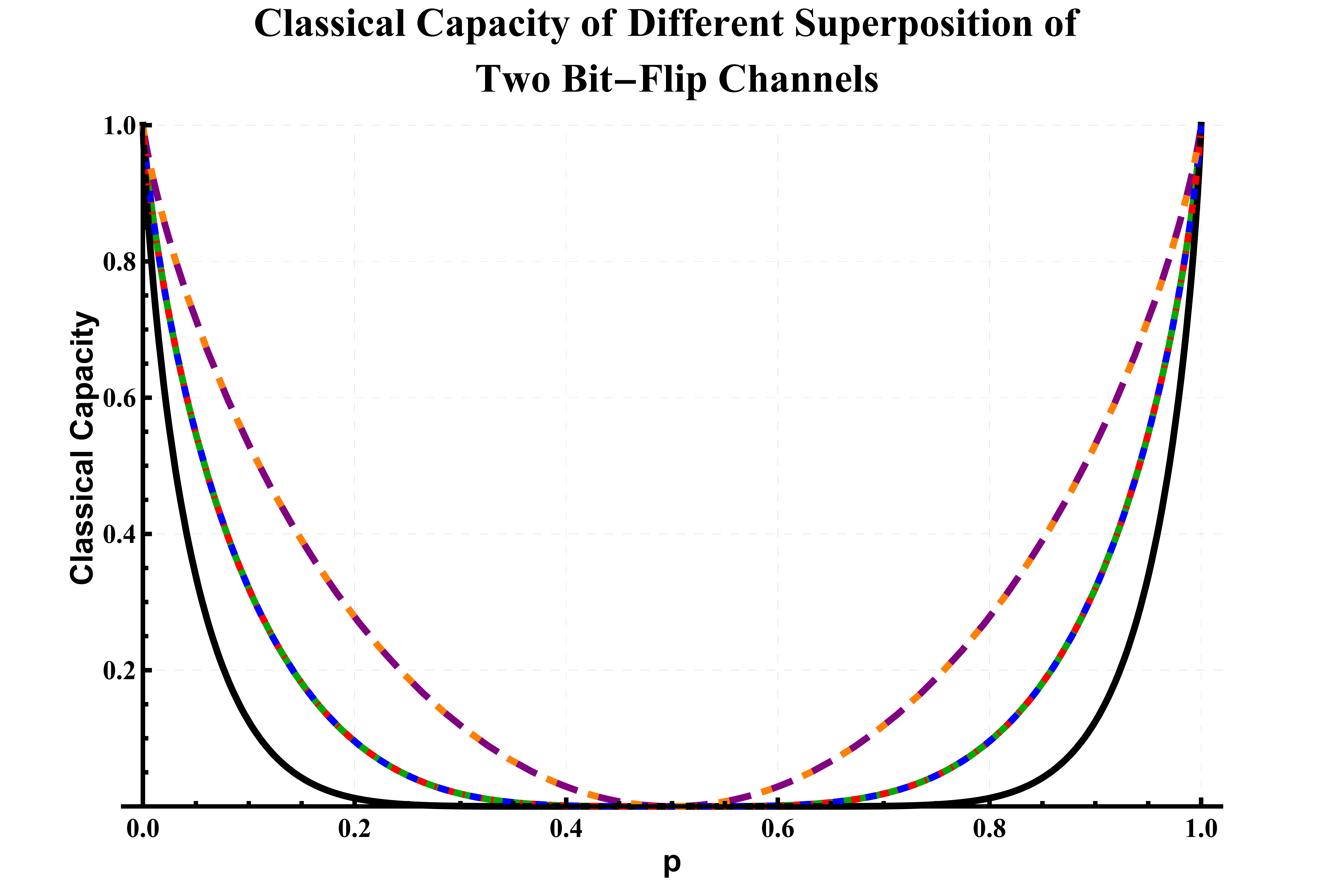}}}
    \subfloat[\label{fig:classical02}]{\fbox{\includegraphics[width=0.32\textwidth]{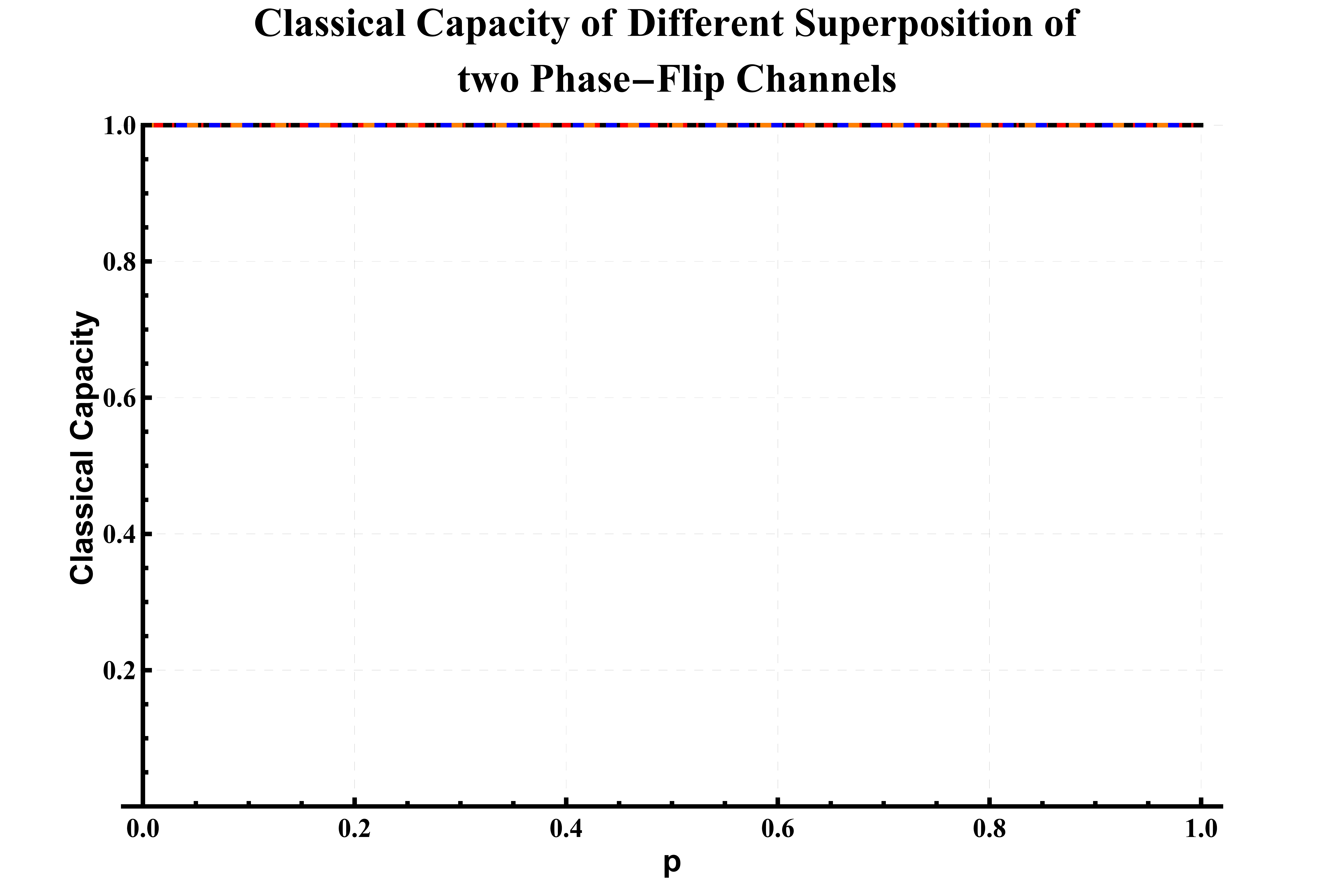}}}
    \subfloat[\label{fig:classical03}]{\fbox{\includegraphics[width=0.32\textwidth]{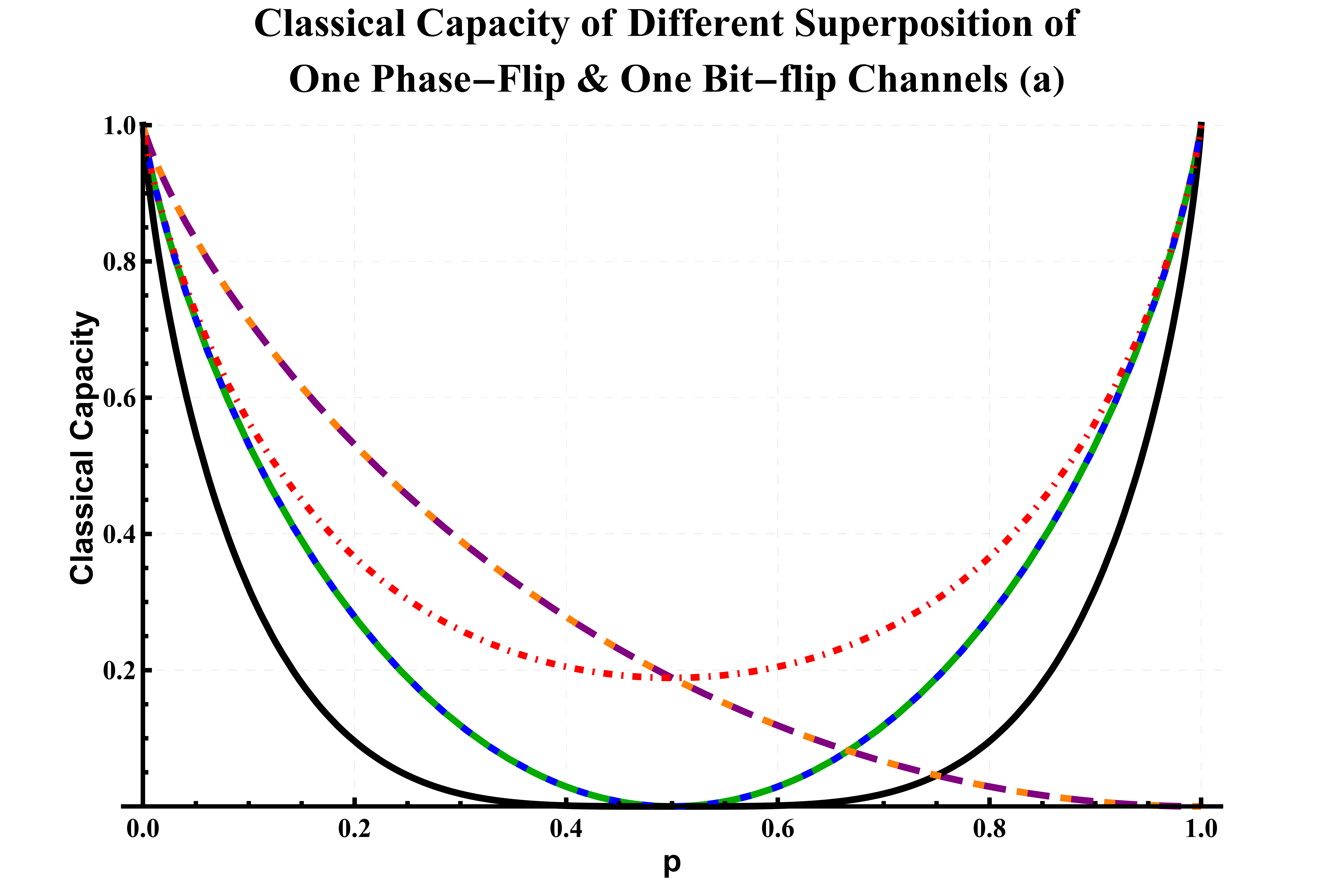}}}

    \subfloat[\label{fig:classical04}]{\fbox{\includegraphics[width=0.32\textwidth]{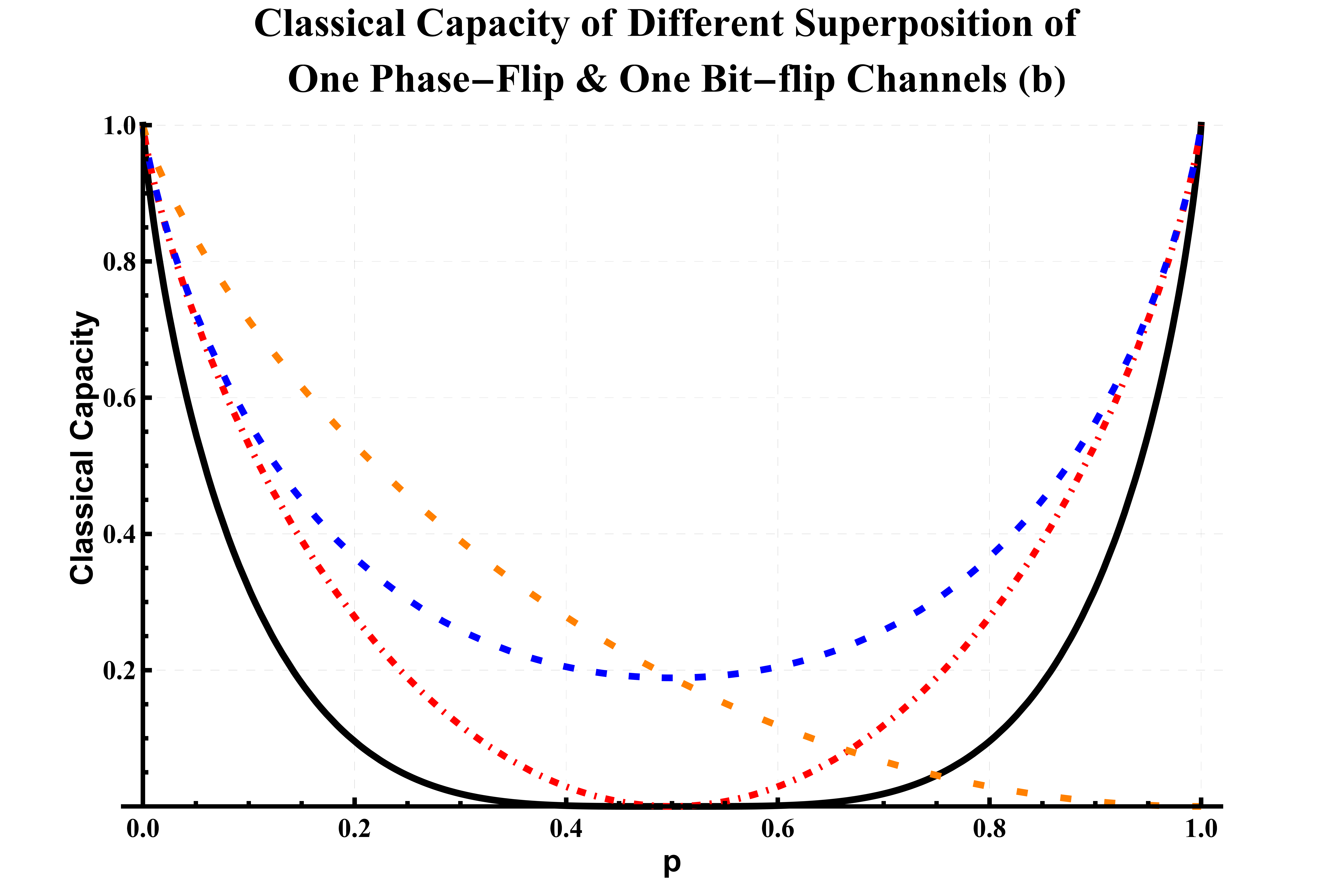}}}
    \subfloat[\label{fig:classical05}]{\fbox{\includegraphics[width=0.32\textwidth]{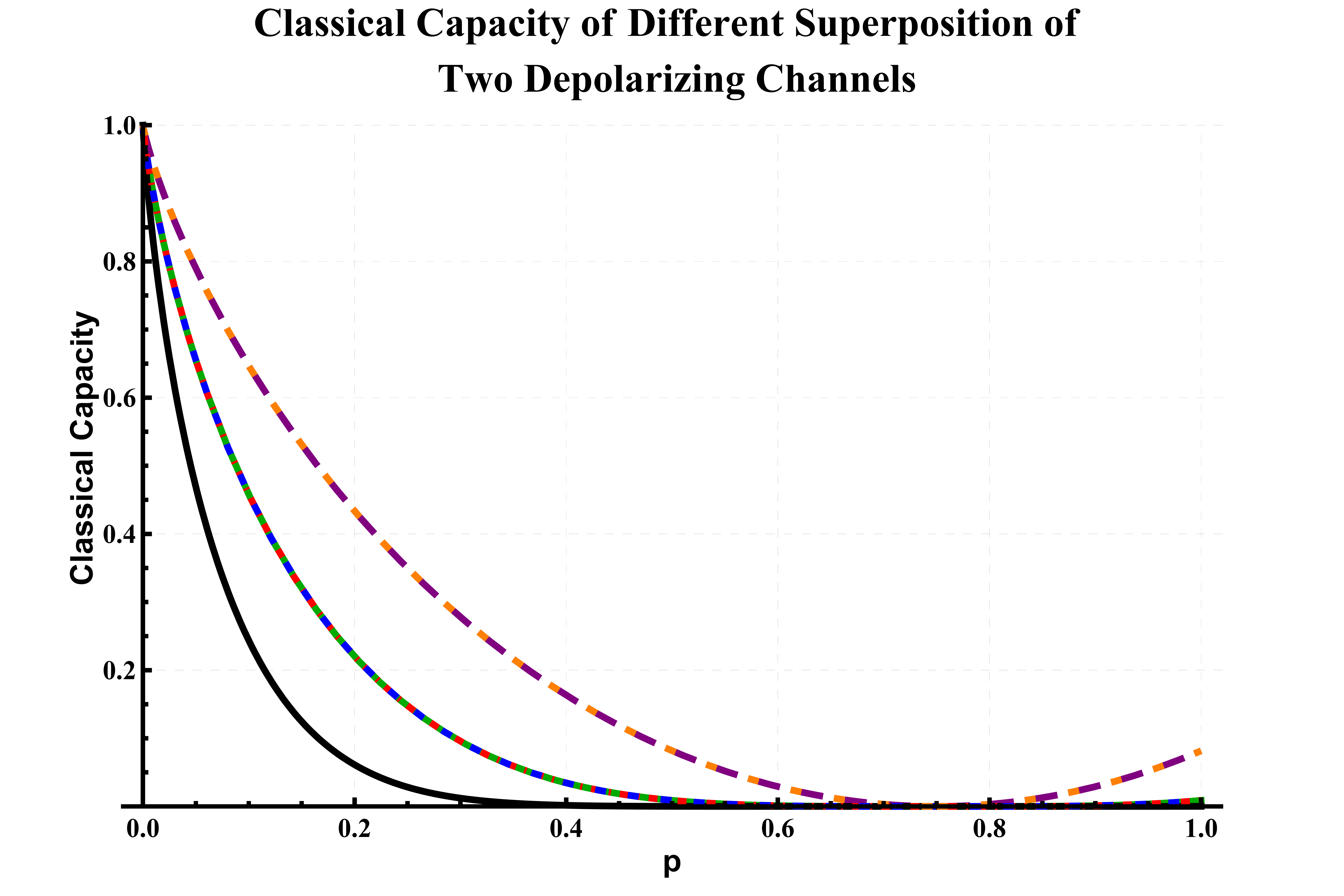}}}
    \hfill\hfill
    \resizebox{0.28\textwidth}{!}{\plotlegend}
    \hfill
    
    \caption{Contour plots of Holevo information evaluated using the coherent control framework for different channel combinations: (a)  Two bit-flip channels, (b) Two phase-flip channels, and (c) One phase flip and another is a bit-flip in the combination $\mathcal{E}_{bit}-\mathcal{E}_{phase}-\mathcal{E}_{bit}-\mathcal{E}_{phase}$ channels. (d)One phase flip and another is a bit-flip in the combination $\mathcal{E}_{bit}-\mathcal{E}_{bit}-\mathcal{E}_{phase}-\mathcal{E}_{phase}$ channels $\&$(e) Two-phase flip channels}
    \label{fig:Classical_Capacity}
\end{figure*}

The graph shown in Fig.~\ref{fig:classical01} is the classical capacity of different superpositions of Bit-Flip channels, where each bit-flip channel noise parameter is $p$. The graph shown in Fig.~\ref{fig:classical02} is the classical capacity of different superpositions of phase-flip channels, where each Phase-flip channel's noise parameter is $p$. 
The classical capacity for the quantum switch configuration with bit-flip and phase-flip channels, as well as their coherent superposition, indicated by the green and purple lines, respectively, is shown in Fig.~\ref{fig:classical03}. The quantum switch of quantum switch (SoS) $\mathcal{S}^{\prime}  \left( \mathcal{S}_1, \mathcal{S}_2  \right)$, which consists of two switches, $\mathcal{S}_1  \left(   \mathcal{E}^{(1)}, \mathcal{E}^{(2)} \right)$ and $\mathcal{S}_2  \left(  \mathcal{E}^{(3)}, \mathcal{E}^{(4)} \right)$, If the channels $\mathcal{E}^{(1)} \& \mathcal{E}^{(3)}$ are bit-flip channels and $\mathcal{E}^{(2)} \& \mathcal{E}^{(4)}$ are phase-flip channels, can be represented as $\mathcal{E}_{bit}-\mathcal{E}_{phase}-\mathcal{E}_{bit}-\mathcal{E}_{phase}$, then the classical capacity of this supermap is represented in Fig.~\ref{fig:classical03}. Similarly, if the combination of bit-flip and phase flip channels is in the $\mathcal{E}_{bit}-\mathcal{E}_{bit}-\mathcal{E}_{phase}-\mathcal{E}_{phase}$ way, then the classical capacities of the mentioned supermaps are shown in the  Fig.~\ref{fig:classical04}. And if each supermap has the $\mathcal{E}^{(i)}$ as depolarize channels, where $\{i\}_1^3$ is from, then the capacities of different superpositions are shown in the Fig.~\ref{fig:classical05}.

\subsection{Quantum Capacities of Different Channels}

\begin{figure*}[tp]
    \centering
    \subfloat[\label{fig:quantum01}]{\fbox{\includegraphics[width=0.32\textwidth]{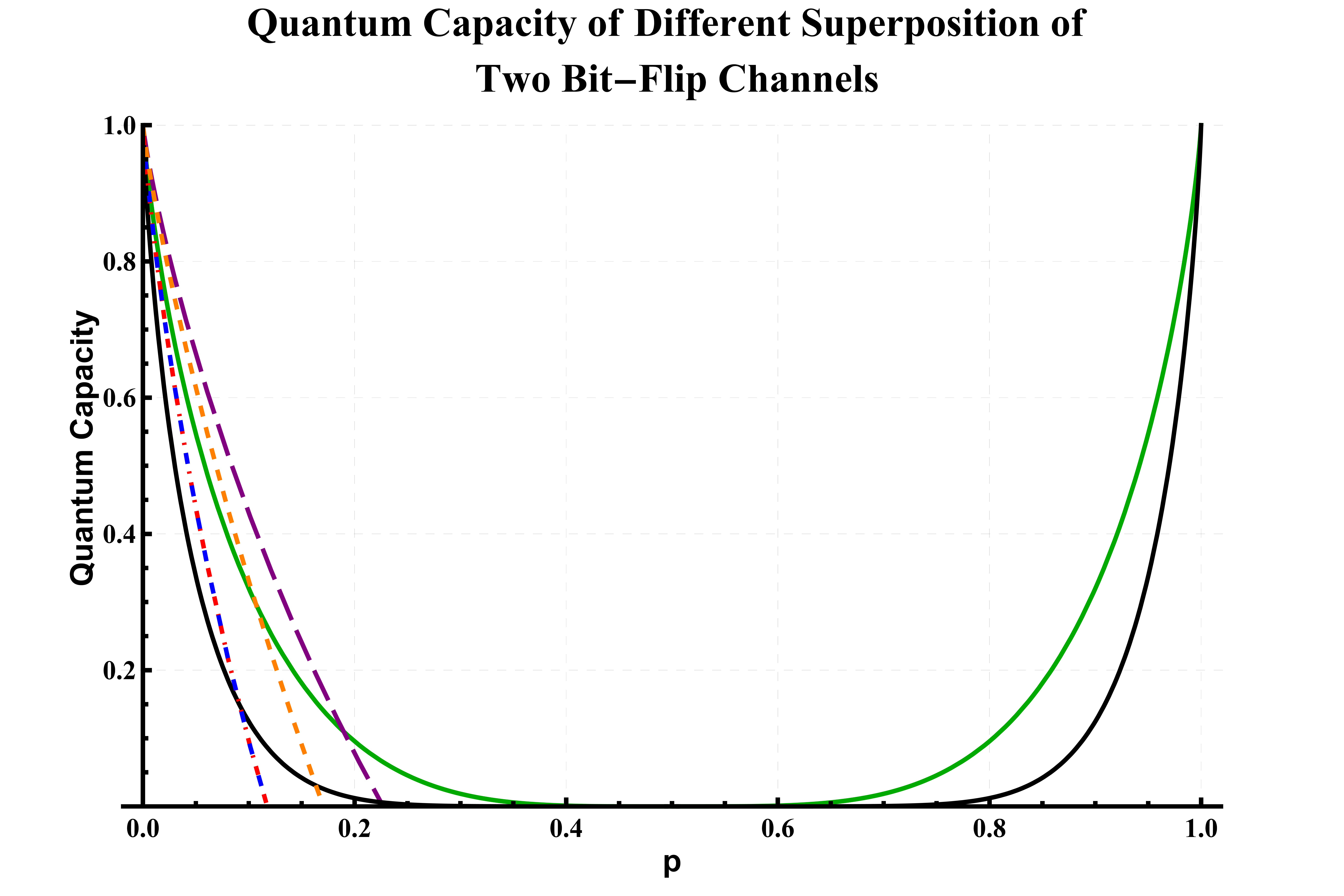}}}
    \subfloat[\label{fig:quantum02}]{\fbox{\includegraphics[width=0.32\textwidth]{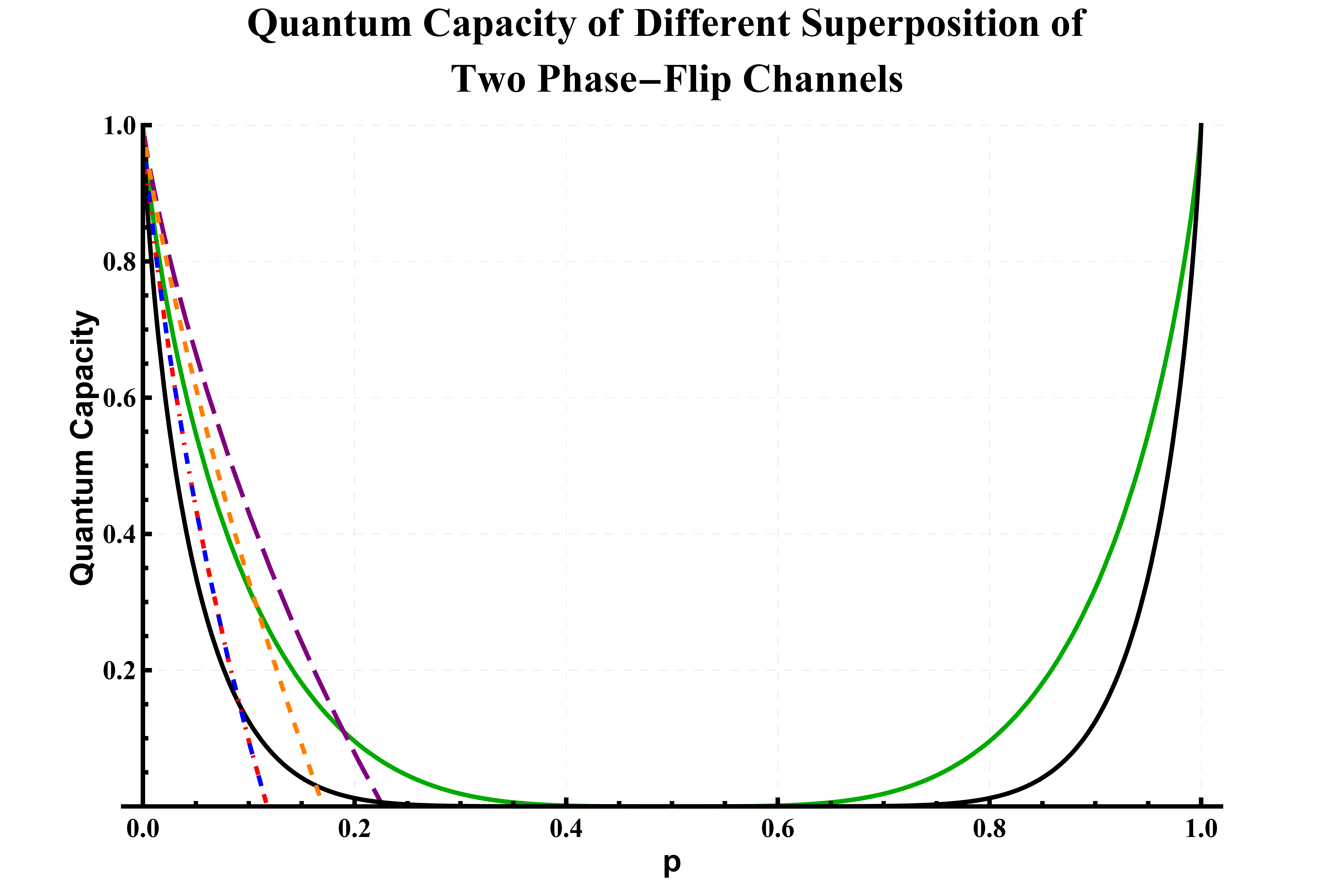}}}
    \subfloat[\label{fig:quantum03}]{\fbox{\includegraphics[width=0.32\textwidth]{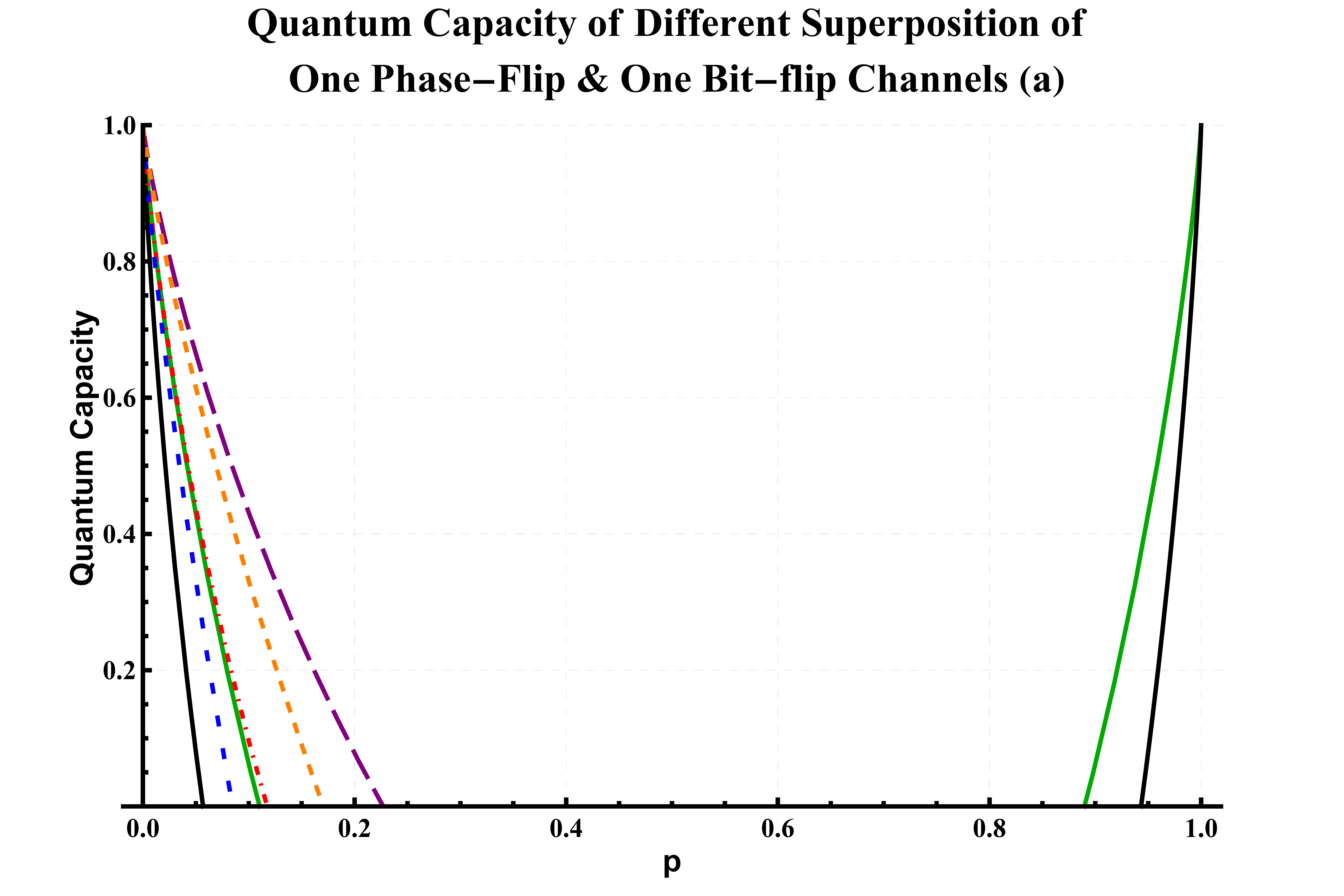}}}

    \subfloat[\label{fig:quantum04}]{\fbox{\includegraphics[width=0.32\textwidth]{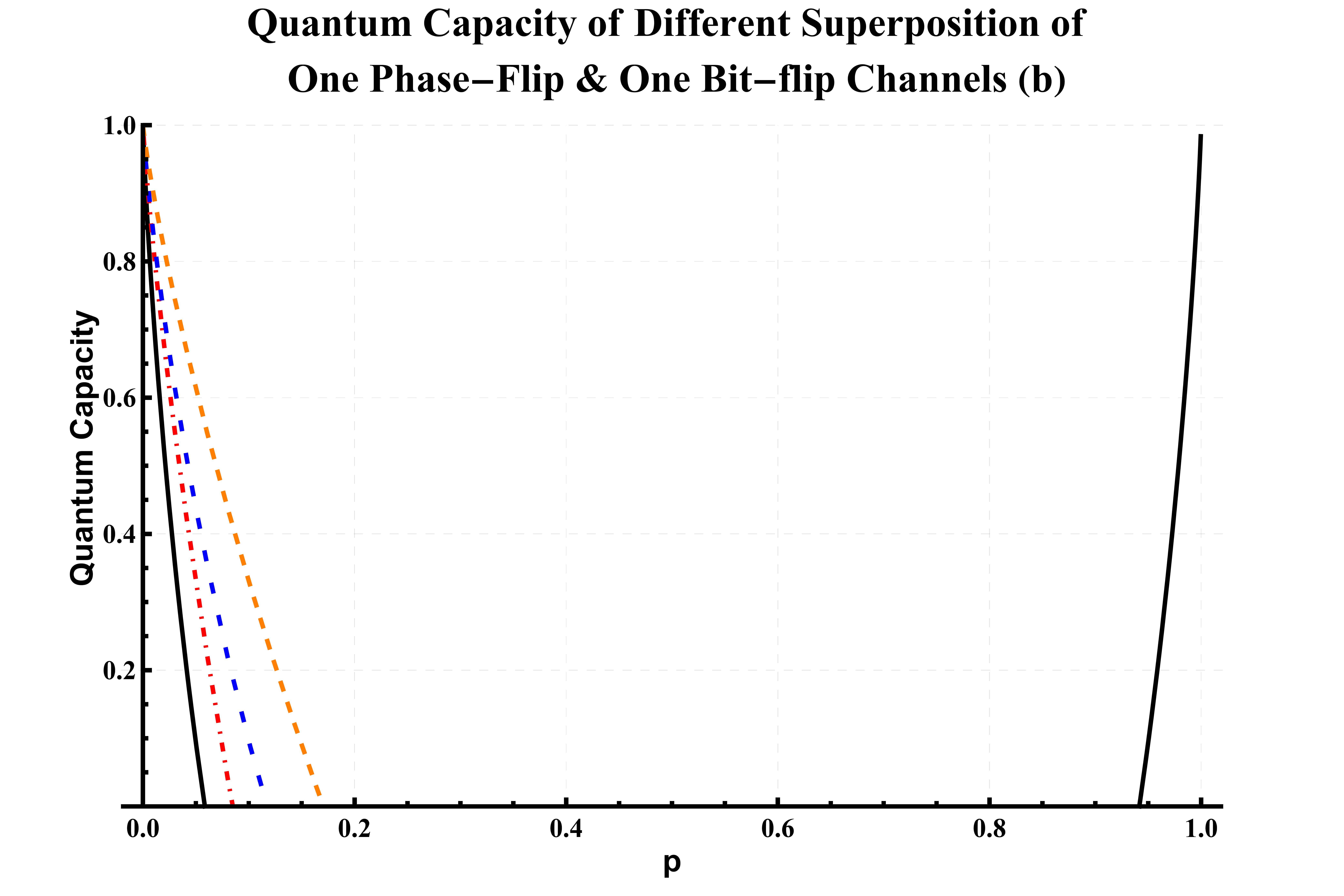}}}
    \subfloat[\label{fig:quantum05}]{\fbox{\includegraphics[width=0.32\textwidth]{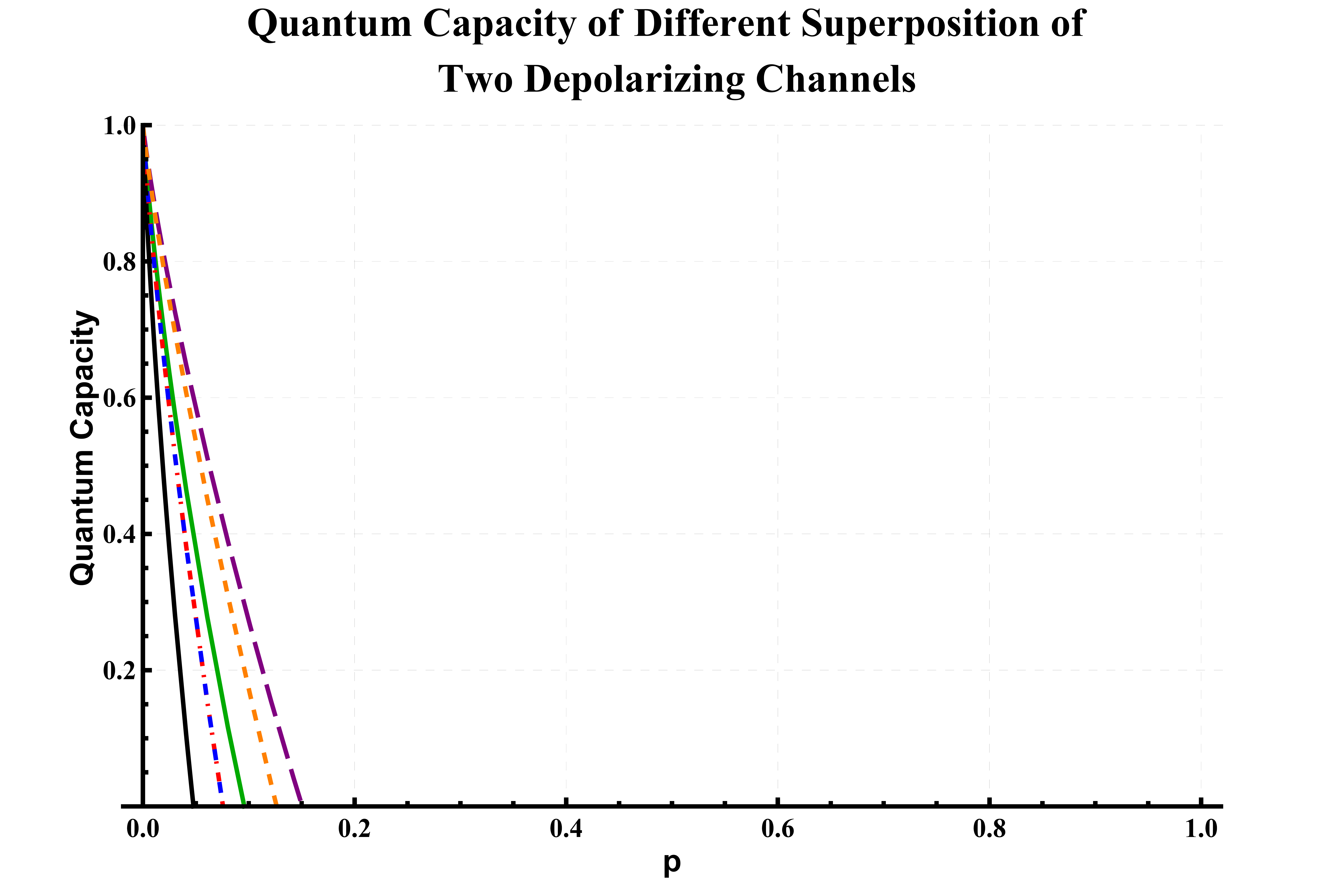}}}
    \hfill\hfill
    \resizebox{0.28\textwidth}{!}{\plotlegend}
    \hfill
    
    \caption{Contour plots of coherent information evaluated using the coherent control framework for different channel combinations: (a)  Two bit-flip channels, (b) Two phase-flip channels, and (c) One phase flip and another is a bit-flip in the combination $\mathcal{E}_{bit}-\mathcal{E}_{phase}-\mathcal{E}_{bit}-\mathcal{E}_{phase}$ channels. (d)One phase flip and another is a bit-flip in the combination $\mathcal{E}_{bit}-\mathcal{E}_{bit}-\mathcal{E}_{phase}-\mathcal{E}_{phase}$ channels $\&$ (e) Two-phase flip channels}
    \label{fig:Quantum_Capacity}
\end{figure*}

Fig.~\ref{fig:quantum01} shows the quantum capacity for the superposition of bit-flip channels. Similarly, Fig.~\ref{fig:quantum02} is for superposition for phase flip channels. Fig.~\ref{fig:quantum03} is the superposition of one bit-flip and one phase flip channel for a quantum switch and coherent superposition, and the combination $\mathcal{E}_{bit}-\mathcal{E}_{phase}-\mathcal{E}_{bit}-\mathcal{E}_{phase}$. Fig.~\ref{fig:quantum04} for the different superposition combination of $\mathcal{E}_{bit}-\mathcal{E}_{bit}-\mathcal{E}_{phase}-\mathcal{E}_{phase}$ channel. And the last Fig.~\ref{fig:quantum05} illustrates the quantum capacity of different superpositions of a depolarizing channel.

\section{Discussion}\label{Disuss}

The study \citet{ebler2018enhanced} found that by sending classical information through a quantum switch of two completely depolarizing channels, which are zero-capacity channels, the overall switch has non-zero capacity. So, the superposition of quantum channels has great advantage in communication. Two important superpositions are the quantum switch and coherent information, which have been appropriately studied. In this manuscript, we analyze the capacity of the combination of these higher-order maps, whose capacities are mentioned in the Fig.~\ref{fig:Classical_Capacity} $\&$ Fig.~\ref{fig:Quantum_Capacity}.

The classical capacities of coherent superposition two bit-flip channels and coherent superposition of coherent superposition of four bit-flip channels are the same. Similarly, this behavior is the same for phase-flip channels, depolarizing channels, or any other kind of combination. So, coherent superposition of two channels, $\boldsymbol{\mathcal{E}^{(1)}}$ $\&$ $\boldsymbol{\mathcal{E}^{(2)}}$ has same capacity of coherent superposition of coherent superposition of four channels, $\boldsymbol{\mathcal{E}^{(1)}}$, $\boldsymbol{\mathcal{E}^{(2)}}$,$\boldsymbol{\mathcal{E}^{(3)}}$, $\&$ $\boldsymbol{\mathcal{E}^{(4)}}$. So, this has no advantages in this supermap compared to a single coherent superposition, in the case of sending classical information through this.

A quantum switch, a quantum switch of coherent superposition (SoC) $\&$ coherent superposition of quantum switch (CoS) have same classical capacity, if all the channels are identical. If the channels are in the combination (a) $\mathcal{E}_{bit}-\mathcal{E}_{phase}-\mathcal{E}_{bit}-\mathcal{E}_{phase}$, then the SoC has more capacity than any other combination. Similarly for the combination $\mathcal{E}_{bit}-\mathcal{E}_{bit}-\mathcal{E}_{phase}-\mathcal{E}_{phase}$, CoS has more classical capacity than any other superpositions. And these two are the same.

But the classical capacity of a quantum switch of quantum switch (SoS) is reduced. And it is less than any other superposition, including one from a single quantum switch, except in one case, where all channels are phase-flip channels. If all the channels are phase flip channels, then the classical capacity of any combination is equal and has value one at any value of $p$. These combinations are perfect for any channel that has only bit-flip errors. 

The most general or depolarizing channel is a simplified version of a Pauli channel. The most efficient way to send classical information is coherent superposition of channel $\&$ CoC. Because both have the same classical capacity, which is greater than any other superposition. So, there is no advantage for a depolarizing channel for SoS, Soc, CoS $\&$ CoC. The most efficient way is coherent superposition of two depolarizing channels.

In conclusion, if there is one kind of channel, such as a bit-flip, phase-flip, or depolarizing channel, the most efficient usage to transmit classical information is coherent superposition of two identical channels. Coherent superposition of coherent superposition also has the same classical capacity, but it needs four channels. However, the same classical capacity can be achieved by coherent superposition, so this is the most efficient method. But if there are two types of channel phase flip-channel and a bit-flip channel, then the most efficient way to send information is either quantum switch of coherent information in the combination $\mathcal{E}_{bit}-\mathcal{E}_{phase}-\mathcal{E}_{bit}-\mathcal{E}_{phase}$, or coherent superposition of quantum switch for the combination $\mathcal{E}_{bit}-\mathcal{E}_{bit}-\mathcal{E}_{phase}-\mathcal{E}_{phase}$.

The quantum capacity remains same for any superposition of bit-flip channels and phase-flip channels. For example, the quantum capacity of a quantum switch of two bit-flip channels has the same quantum capacity as a quantum switch of two phase-flip channels, and so on. The quantum switch and quantum switch of switch (SoS) have high quantum capacity when the noise parameter $p$ is almost 1 for a bit-flip and phase-flip channel or any combination of bit-flip and phase-flip channels. However, for these cases, a quantum switch has more quantum capacity than. The SoS also does not have any advantages like its classical capacity. For the combination, $\mathcal{E}_{bit}-\mathcal{E}_{phase}-\mathcal{E}_{bit}-\mathcal{E}_{phase}$, at a low value of $p$, CoC has more capacity than any other superposition technique. And, also for the combination$\mathcal{E}_{bit}-\mathcal{E}_{bit}-\mathcal{E}_{phase}-\mathcal{E}_{phase}$, CoC has more capacity, but it is less than the quantum capacity of a coherent superposition of one bit-flip channel and one-phase flip channel.

Also, for the superpositions of the depolarizing channel, the coherent superposition of two depolarizing channels has more quantum capacity than any other superposition.

\section{Conclusion}
In this work, we have conducted an extensive analysis of various compositions of quantum channels realized through quantum switch and path superposition configurations. We evaluated their single-shot classical and quantum capacities and performed a detailed comparative study.

Our findings reveal several noteworthy trends. A quantum switch composed of two quantum switches exhibits a lower capacity than a single switch. Similarly, a coherent superposition of two coherent superpositions yields the same capacity as a single coherent superposition. Furthermore, both the quantum switch of a coherent superposition and the coherent superposition of a quantum switch exhibit nearly identical capacities to that of a single quantum switch, which, in certain cases, is lower than that of a coherent superposition of two channels.

Overall, the coherent superposition of two quantum channels demonstrates superior performance in specific scenarios, emerging as the most efficient framework for transmitting both classical and quantum information through a noisy quantum channel.

\textbf{Acknowledgements-}The authors would like to thank Rajiuddin Sk for fruitful discussions and also thank Dr. Debarshi Das for his assistance in understanding and helping with codes. Additionally, they appreciate a helpful conversation with Prof. Giulio Chiribella.

\bibliographystyle{apsrev4-2}
\bibliography{Reference}

\appendix
\section{Expressions for classical capacity of different superpositions}

\subsection{Quantum Switch}
The classical capacity of a quantum switch of two bit-flip channels, two phase-flip channels, one bit-flip and phase-flip channel, and of two depolarizing channels is given, respectively.

\begin{align*}
C_{\text{switch}}^{1} &= \frac{1}{\log(2)} \Biggl( 4p(p-1) \operatorname{ArcTanh}\left((1 - 2p)^2\right) \\
& \qquad + \log\left(2 + 4p(p-1)\right) \Biggr) \\[1em]
C_{\text{switch}}^{2} &= 1 \\[1em]
C_{\text{switch}}^{3} &= \frac{\log(2 - 2p) + p \log\left(\frac{p}{1-p}\right)}{\log(2)} \\[1em]
C_{\text{switch}}^{4} &= \frac{1}{\log(512)} \Biggl( 8 p (2p-3) \operatorname{arccoth}\left(\frac{9}{(3 - 4 p)^2}\right) \\
& \qquad + \log(512) \\
& \qquad + 9 \log\left(1 + \frac{4}{9} p (2p-3)\right) \Biggr)
\end{align*}

\subsection{Coherent Superposition}

The classical capacity of a coherent superposition of two bit-flip channels, two phase-flip channels, one bit-flip and phase-flip channel, and of two depolarizing channels is given, respectively.

\begin{align*}
C_{\text{coh}}^{1} &= \frac{\log(2 - 2p) + p \log\left(\frac{p}{1-p}\right)}{\log(2)} \\[1em]
C_{\text{coh}}^{2} &= 1 \\[1em]
C_{\text{coh}}^{3} &= \frac{(2 - p) \log(2 - p) + p \log(p)}{\log(4)} \\[1em]
C_{\text{coh}}^{4} &= 1 - \frac{4p \operatorname{ArcTanh}\left(1 - \frac{4p}{3}\right)}{\log(8)} \\
& \qquad + \frac{3 \log\left(1 - \frac{2p}{3}\right)}{\log(8)}
\end{align*}

\subsection{SoS}

Classical capacity of a quantum switch of quantum switch (SoS) two bit-flip channels, two phase-flip channels, one bit-flip and phase-flip channel in the combination $\mathcal{E}_{bit}-\mathcal{E}_{phase}-\mathcal{E}_{bit}-\mathcal{E}_{phase}$,  one bit-flip and phase-flip channel in the combination $\mathcal{E}_{bit}-\mathcal{E}_{bit}-\mathcal{E}_{phase}-\mathcal{E}_{phase}$, and two depolarizing channels, respectably

\begin{align*}
C_{\text{CoS}}^{1} &= \frac{1}{\log(2)} \Biggl( 4p(p-1) \operatorname{ArcTanh}\left((1 - 2p)^2\right) \\
& \qquad + \log\left(2 + 4p(p-1)\right) \Biggr) \\[1em]
C_{\text{SoC}}^{2} &= 1 \\[1em]
C_{\text{CoS}}^{3} &= 1 + p\log_2(p) + (1-p)\log_2(1-p) \\[1em]
C_{\text{CoS}}^{4} &= 1 + \log_2(1 + p(p-1)) \\
& \qquad + \frac{2p(p-1) \operatorname{ArcTanh}\left(1 + 2p(p-1)\right)}{\log(2)} \\[1em]
C_{\text{CoS}}^{5} &= 1 + \frac{8p(2p-3) \text{arccoth}\left(\frac{9}{(3 - 4 p)^2}\right)}{9\log(2)} \\
& \qquad + \log_2\left(1 + \frac{4}{9} p (2p - 3)\right)
\end{align*}

\subsection{SoC}

Classical capacity of a quantum switch of a quantum coherent superposition (SoC) of two bit-flip channel, two phase-flip channels, one bit-flip and phase-flip channel in the combination $\mathcal{E}_{bit}-\mathcal{E}_{phase}-\mathcal{E}_{bit}-\mathcal{E}_{phase}$,  one bit-flip and phase-flip channel in the combination $\mathcal{E}_{bit}-\mathcal{E}_{bit}-\mathcal{E}_{phase}-\mathcal{E}_{phase}$, and two depolarizing channels, respectably

\begin{align*}
C_{\text{SoC}}^{1} &= \frac{1}{\log(2)} \Biggl( 4p(p-1) \operatorname{ArcTanh}\left((1 - 2p)^2\right) \\
& \qquad + \log\left(2 + 4p(p-1)\right) \Biggr) \\[1em]
C_{\text{SoC}}^{2} &= 1 \\[1em]
C_{\text{SoC}}^{3} &= \frac{1}{\log(2)} \Biggl( 2p(p-1) \operatorname{ArcTanh}\left(1 + 2p(p-1)\right) \\
& \qquad + \log\left(2 + 2p(p-1)\right) \Biggr) \\[1em]
C_{\text{SoC}}^{4} &= 1 + p\log_2(p) + (1-p)\log_2(1-p) \\[1em]
C_{\text{SoC}}^{5} &= 1 + \frac{8p(2p-3) \text{arccoth}\left(\frac{9}{(3 - 4 p)^2}\right)}{9\log(2)} \\
& \qquad + \log_2\left(1 + \frac{4}{9} p (2p - 3)\right)
\end{align*}

\subsection{CoS}

Classical capacity of a coherent superposition of a quantum switch (CoS) of two two-bit-flip channel, two phase-flip channels, one bit-flip and phase-flip channel in the combination $\mathcal{E}_{bit}-\mathcal{E}_{phase}-\mathcal{E}_{bit}-\mathcal{E}_{phase}$,  one bit-flip and phase-flip channel in the combination $\mathcal{E}_{bit}-\mathcal{E}_{bit}-\mathcal{E}_{phase}-\mathcal{E}_{phase}$, and two depolarizing channels, respectably

\begin{align*}
C_{\text{CoS}}^{1} &= \frac{1}{\log(2)} \Biggl( 4p(p-1) \operatorname{ArcTanh}\left((1 - 2p)^2\right) \\
& \qquad + \log\left(2 + 4p(p-1)\right) \Biggr) \\[1em]
C_{\text{CoS}}^{2} &= 1 \\[1em]
C_{\text{CoS}}^{3} &= 1 + p\log_2(p) + (1-p)\log_2(1-p) \\[1em]
C_{\text{CoS}}^{4} &= 1 + \log_2(1 + p(p-1)) \\
& \qquad + \frac{2p(p-1) \operatorname{ArcTanh}\left(1 + 2p(p-1)\right)}{\log(2)} \\[1em]
C_{\text{CoS}}^{5} &= 1 + \frac{8p(2p-3) \text{arccoth}\left(\frac{9}{(3 - 4 p)^2}\right)}{9\log(2)} \\
& \qquad + \log_2\left(1 + \frac{4}{9} p (2p - 3)\right)
\end{align*}

\subsection{CoC}

Classical capacity of a coherent superposition of coherent superposition (CoC) of two two-bit-flip channel, two phase-flip channels, one bit-flip and phase-flip channel in the combination $\mathcal{E}_{bit}-\mathcal{E}_{phase}-\mathcal{E}_{bit}-\mathcal{E}_{phase}$,  one bit-flip and phase-flip channel in the combination $\mathcal{E}_{bit}-\mathcal{E}_{bit}-\mathcal{E}_{phase}-\mathcal{E}_{phase}$, and two depolarizing channels, respectably

\begin{align*}
C_{\text{CoC}}^{1} &= 1 + p\log_2(p) + (1-p)\log_2(1-p) \\[1em]
C_{\text{CoC}}^{2} &= 1 \\[1em]
C_{\text{CoC}}^{3} &= \frac{1}{2} \left[ (2-p)\log_2(2-p) + p\log_2(p) \right] \\[1em]
C_{\text{CoC}}^{4} &= \frac{1}{2} \left[ (2-p)\log_2(2-p) + p\log_2(p) \right]\\[1em]
C_{\text{CoC}}^{5} &= 1 - \frac{4p \operatorname{ArcTanh}\left(1 - \frac{4p}{3}\right)}{\log(8)} \\
& \qquad + \frac{3 \log\left(1 - \frac{2p}{3}\right)}{\log(8)}
\end{align*}

\section{Expressions for Quantum capacity of different superpositions}

The expressions for quantum capacity are complicated to write down here, and some expressions are big expressions. For example, the quantum capacity of a quantum switch of two bit-flip channels is

\begin{align*}
Q_{\text{switch}}^{1} ={} & 1 \\
& - 2.8853900817779268 \, p(p-1) \\
& \qquad \times \log\left(-2p(p-1)\right) \\
& + 1.4426950408889634 \left(1 + 2p(p-1)\right) \\
& \qquad \times \log\left(1 + 2p(p-1)\right)
\end{align*}

\end{document}

%% file: diagram_switch.tex
\newcommand{\diagramswitch}{%
\resizebox{\columnwidth}{!}{%
  \begin{tikzpicture}[
	thick,
	gate/.style={draw, fill=gray!20, minimum height=0.6cm, minimum width=0.6cm},
	midarrow/.style={
		decoration={
			markings,
			mark=at position 0.5 with {\arrow{stealth}}
		},
		postaction={decorate}
	},
	curvearrow/.style={
		decoration={
			markings,
			mark=at position 0.7 with {\arrow{stealth}}
		},
		postaction={decorate}
	}
	]
	
	\node (rho_c) at (0, 1.5) {$\rho_c $};
	\node (rho_t) at (0, 0)    {$\rho_t$};
	
	\coordinate (control_source) at (1, 1.5); 
	\coordinate (control_target) at (1, 0);   
	
	\coordinate (fork)   at (1.5, 0);   
	\coordinate (rejoin) at (4.5, 0);   
	
	\node (rho_out) at (6, 0)   {$\rho_{out}$};
	
	\node[gate] (gate_A) at (3, 0.75) {$\boldsymbol{\mathcal{E}}^{(1)}$};
	\node[gate] (gate_B) at (3, -0.75) {$\boldsymbol{\mathcal{E}}^{(2)}$};
	
	\draw (rho_c.east) -- (control_source) -- (6, 1.5);
	\draw (control_source) -- (control_target);
	
	\draw[midarrow] (rho_t.east) -- (control_target);
	\draw[-stealth] (control_target) -- (fork);     
	\draw[midarrow] (fork) |- (gate_A.west);
	\draw[midarrow] (gate_A.east) -| (rejoin);
	\draw[midarrow] (fork) |- (gate_B.west);
	\draw[midarrow] (gate_B.east) -| (rejoin);
	\draw[midarrow] (rejoin) -- (rho_out.west);
	
	\draw[curvearrow, red] (gate_A.east) .. controls (4.5, 0.4) and (1.5, -0.4) .. (gate_B.west);
	\draw[curvearrow, green!70!black] (gate_B.east) .. controls (4.5, -0.4) and (1.5, 0.4) .. (gate_A.west);
	
	\filldraw (control_source) circle (1.5pt); % Solid circle (back to original size)
	\draw (control_target) circle (2.5pt);      % Hollow circle (remains large)
	
\end{tikzpicture}%
}%
}

\newcommand{\diagramswitchaa}{%
\resizebox{\columnwidth}{!}{%
  \begin{tikzpicture}[
	thick,
	gate/.style={draw, fill=gray!20, minimum height=0.6cm, minimum width=0.6cm},
	midarrow/.style={
		decoration={
			markings,
			mark=at position 0.5 with {\arrow{stealth}}
		},
		postaction={decorate}
	},
	curvearrow/.style={
		decoration={
			markings,
			mark=at position 0.7 with {\arrow{stealth}}
		},
		postaction={decorate}
	}
	]
	
	\node (rho_c) at (0, 1.5) {$\dyad{0}{0}$};
	\node (rho_t) at (0, 0)    {$\rho_t$};
	
	\coordinate (control_source) at (1, 1.5); 
	\coordinate (control_target) at (1, 0);   
	
	\coordinate (fork)   at (1.5, 0);   
	\coordinate (rejoin) at (4.5, 0);   
	
	\node (rho_out) at (6, 0)   {$\rho_{out}$};
	
	\node[gate] (gate_A) at (3, 0.75) {$\boldsymbol{\mathcal{E}}^{(1)}$};
	\node[gate] (gate_B) at (3, -0.75) {$\boldsymbol{\mathcal{E}}^{(2)}$};
	
	\draw (rho_c.east) -- (control_source) -- (6, 1.5);
	\draw (control_source) -- (control_target);
	
	\draw[midarrow] (rho_t.east) -- (control_target);
	\draw[-stealth] (control_target) -- (fork);     
	\draw[midarrow] (fork) |- (gate_A.west);
	
	\draw[midarrow] (gate_B.east) -| (rejoin);
	\draw[midarrow] (rejoin) -- (rho_out.west);
	
	\draw[curvearrow, red] (gate_A.east) .. controls (4.5, 0.4) and (1.5, -0.4) .. (gate_B.west);

	\filldraw (control_source) circle (1.5pt); % Solid circle (back to original size)
	\draw (control_target) circle (2.5pt);      % Hollow circle (remains large)
	
\end{tikzpicture}%
}%
}

\newcommand{\diagramswitchbb}{%
\resizebox{\columnwidth}{!}{%
  \begin{tikzpicture}[
	thick,
	gate/.style={draw, fill=gray!20, minimum height=0.6cm, minimum width=0.6cm},
	midarrow/.style={
		decoration={
			markings,
			mark=at position 0.5 with {\arrow{stealth}}
		},
		postaction={decorate}
	},
	curvearrow/.style={
		decoration={
			markings,
			mark=at position 0.7 with {\arrow{stealth}}
		},
		postaction={decorate}
	}
	]
	
	\node (rho_c) at (0, 1.5) {$\dyad{1}{1}$};
	\node (rho_t) at (0, 0)    {$\rho_t$};
	
	\coordinate (control_source) at (1, 1.5); 
	\coordinate (control_target) at (1, 0);   
	
	\coordinate (fork)   at (1.5, 0);   
	\coordinate (rejoin) at (4.5, 0);   
	
	\node (rho_out) at (6, 0)   {$\rho_{out}$};
	
	\node[gate] (gate_A) at (3, 0.75) {$\boldsymbol{\mathcal{E}}^{(1)}$};
	\node[gate] (gate_B) at (3, -0.75) {$\boldsymbol{\mathcal{E}}^{(2)}$};
	
	\draw (rho_c.east) -- (control_source) -- (6, 1.5);
	\draw (control_source) -- (control_target);
	
	\draw[midarrow] (rho_t.east) -- (control_target);
	\draw[-stealth] (control_target) -- (fork);     
	
	\draw[midarrow] (gate_A.east) -| (rejoin);
	\draw[midarrow] (fork) |- (gate_B.west);

	\draw[midarrow] (rejoin) -- (rho_out.west);
	
	\draw[curvearrow, green!70!black] (gate_B.east) .. controls (4.5, -0.4) and (1.5, 0.4) .. (gate_A.west);
	
	\filldraw (control_source) circle (1.5pt); % Solid circle (back to original size)
	\draw (control_target) circle (2.5pt);      % Hollow circle (remains large)
	
\end{tikzpicture}%
}%
}

%% file: diagram_superposition.tex
\newcommand{\diagramcoherent}{%
    \begin{tikzpicture}[
	thick,
	gate/.style={draw, fill=gray!20, minimum height=0.6cm, minimum width=0.6cm},
	midarrow/.style={
		decoration={
			markings,
			mark=at position 0.5 with {\arrow{stealth}}
		},
		postaction={decorate}
	}
	]
	
	\node (rho_c) at (0, 1.5) {$\rho_c$};
	\node (rho_t) at (0, 0)    {$\rho_t$};
	
	\coordinate (control_source) at (1, 1.5); % Point on top wire
	\coordinate (control_target) at (1, 0);   % Point on middle wire
	
	\coordinate (fork)   at (1.5, 0);   % Point where the path splits (shifted right)
	\coordinate (rejoin) at (4.5, 0);   % Point where paths rejoin
	
	\node (rho_out) at (6, 0)   {$\rho_{out}$};
	
	\node[gate] (gate_A) at (3, 0.75) {$\boldsymbol{\mathcal{E}^{(1)}}$};
	\node[gate] (gate_B) at (3, -0.75) {$\boldsymbol{\mathcal{E}^{(2)}}$};
	
	\draw (rho_c.east) -- (control_source) -- (6, 1.5);
	
	\draw (control_source) -- (control_target);
	
	\draw[midarrow] (rho_t.east) -- (control_target);
	\draw[-stealth] (control_target) -- (fork);
	
	\draw[midarrow] (fork) |- (gate_A.west);
	\draw[midarrow] (gate_A.east) -| (rejoin);
	
	\draw[midarrow] (fork) |- (gate_B.west);
	\draw[midarrow] (gate_B.east) -| (rejoin);
	
	\draw[midarrow] (rejoin) -- (rho_out.west);
	
	\filldraw (control_source) circle (1.5pt); % Small solid circle
	\draw (control_target) circle (2.5pt);      % ADDED: Large hollow circle
	
\end{tikzpicture}%
}

%% file: diagram_coc.tex
\newcommand{\diagramcoc}{%
\resizebox{\columnwidth}{!}{%
   \begin{tikzpicture}[
	thick,
	gate/.style={draw, fill=gray!20, minimum height=0.6cm, minimum width=0.6cm},
	midarrow/.style={
		decoration={
			markings,
			mark=at position 0.5 with {\arrow{stealth}}
		},
		postaction={decorate}
	}
]
	
	\node (rho_c) at (0, 3) {$\rho_c$};
	\node (rho_t) at (0, 0) {$\rho_t$};
	\coordinate (main_control_source) at (1, 3);
	\coordinate (main_control_target) at (1, 0);
	\coordinate (main_fork) at (1.5, 0);
	\coordinate (main_rejoin) at (7.5, 0);
	\node (rho_out) at (9, 0) {$\rho_{out}$};

	\pgfmathsetmacro{\yupper}{1.5}
	\coordinate (control_source_upper) at (2.5, 3) {};
	\coordinate (control_target_upper) at (2.5, \yupper) {};
	\coordinate (fork_upper) at (3, \yupper);
	\coordinate (rejoin_upper) at (6, \yupper);
	\node[gate] (gate_A) at (4.5, \yupper+0.75) {$\boldsymbol{\mathcal{E}^{(1)}}$};
	\node[gate] (gate_B) at (4.5, \yupper-0.75) {$\boldsymbol{\mathcal{E}^{(2)}}$};

	\pgfmathsetmacro{\ylower}{-1.5}
	\coordinate (control_target_lower) at (2.5, \ylower) {};
	\coordinate (fork_lower) at (3, \ylower);
	\coordinate (rejoin_lower) at (6, \ylower);
	\node[gate] (gate_C) at (4.5, \ylower+0.75) {$\boldsymbol{\mathcal{E}^{(3)}}$};
	\node[gate] (gate_D) at (4.5, \ylower-0.75) {$\boldsymbol{\mathcal{E}^{(4)}}$};

	\draw (rho_c.east) -- (main_control_source) -- (control_source_upper) -- (9, 3);
	\draw (main_control_source) -- (main_control_target);
	\draw (control_source_upper) -- (control_target_upper);
	\draw (control_source_upper) -- (control_target_lower);

	\draw[midarrow] (rho_t.east) -- (main_control_target);
	\draw[-stealth] (main_control_target) -- (main_fork);
	
	\draw[midarrow] (main_fork) |- (control_target_upper);
	\draw[midarrow] (main_fork) |- (control_target_lower);
	
	\draw[-stealth] (control_target_upper) -- (fork_upper);
	\draw[midarrow] (fork_upper) |- (gate_A.west);
	\draw[midarrow] (gate_A.east) -| (rejoin_upper);
	\draw[midarrow] (fork_upper) |- (gate_B.west);
	\draw[midarrow] (gate_B.east) -| (rejoin_upper);
	\draw[midarrow] (rejoin_upper) -| (main_rejoin);
	
	\draw[-stealth] (control_target_lower) -- (fork_lower);
	\draw[midarrow] (fork_lower) |- (gate_C.west);
	\draw[midarrow] (gate_C.east) -| (rejoin_lower);
	\draw[midarrow] (fork_lower) |- (gate_D.west);
	\draw[midarrow] (gate_D.east) -| (rejoin_lower);
	\draw[midarrow] (rejoin_lower) -| (main_rejoin);
	
	\draw[midarrow] (main_rejoin) -- (rho_out.west);

	\filldraw (main_control_source) circle (1.5pt);
	\filldraw (control_source_upper) circle (1.5pt);
	\draw (main_control_target) circle (2.5pt);
	\draw (control_target_upper) circle (2.5pt);
	\draw (control_target_lower) circle (2.5pt);
	
\end{tikzpicture}%
}%
}

%% file: diagram_cos.tex
\newcommand{\diagramcos}{%
\resizebox{\columnwidth}{!}{%
   \begin{tikzpicture}[
	thick,
	gate/.style={draw, fill=gray!20, minimum height=0.6cm, minimum width=0.6cm},
	midarrow/.style={
		decoration={
			markings,
			mark=at position 0.5 with {\arrow{stealth}}
		},
		postaction={decorate}
	},
	curvearrow/.style={
		decoration={
			markings,
			mark=at position 0.7 with {\arrow{stealth}}
		},
		postaction={decorate}
	}
]
	
	\node (rho_c) at (0, 3) {$\rho_c$};
	\node (rho_t) at (0, 0) {$\rho_t$};
	\coordinate (main_control_source) at (1, 3);
	\coordinate (main_control_target) at (1, 0);
	\coordinate (main_fork) at (1.5, 0);
	\coordinate (main_rejoin) at (7.5, 0);
	\node (rho_out) at (9, 0) {$\rho_{out}$};

	\pgfmathsetmacro{\yupper}{1.5}
	\coordinate (control_source_upper) at (2.5, 3) {};
	\coordinate (control_target_upper) at (2.5, \yupper) {};
	\coordinate (fork_upper) at (3, \yupper);
	\coordinate (rejoin_upper) at (6, \yupper);
	\node[gate] (gate_A) at (4.5, \yupper+0.75) {A};
	\node[gate] (gate_B) at (4.5, \yupper-0.75) {B};

	\pgfmathsetmacro{\ylower}{-1.5}
	\coordinate (control_target_lower) at (2.5, \ylower) {};
	\coordinate (fork_lower) at (3, \ylower);
	\coordinate (rejoin_lower) at (6, \ylower);
	\node[gate] (gate_C) at (4.5, \ylower+0.75) {C};
	\node[gate] (gate_D) at (4.5, \ylower-0.75) {D};

	\draw (rho_c.east) -- (main_control_source) -- (control_source_upper) -- (9, 3);
	\draw (main_control_source) -- (main_control_target);
	\draw (control_source_upper) -- (control_target_upper);
	\draw (control_source_upper) -- (control_target_lower);

	\draw[midarrow] (rho_t.east) -- (main_control_target);
	\draw[-stealth] (main_control_target) -- (main_fork);
	
	\draw[midarrow] (main_fork) |- (control_target_upper);
	\draw[midarrow] (main_fork) |- (control_target_lower);
	
	\draw[-stealth] (control_target_upper) -- (fork_upper);
	\draw[midarrow] (fork_upper) |- (gate_A.west);
	\draw[midarrow] (gate_A.east) -| (rejoin_upper);
	\draw[midarrow] (fork_upper) |- (gate_B.west);
	\draw[midarrow] (gate_B.east) -| (rejoin_upper);
	\draw[midarrow] (rejoin_upper) -| (main_rejoin);
	
	\draw[-stealth] (control_target_lower) -- (fork_lower);
	\draw[midarrow] (fork_lower) |- (gate_C.west);
	\draw[midarrow] (gate_C.east) -| (rejoin_lower);
	\draw[midarrow] (fork_lower) |- (gate_D.west);
	\draw[midarrow] (gate_D.east) -| (rejoin_lower);
	\draw[midarrow] (rejoin_lower) -| (main_rejoin);
	
	\draw[midarrow] (main_rejoin) -- (rho_out.west);

	\draw[curvearrow, red] (gate_A.east) .. controls (5.5, \yupper+0.4) and (3.5, \yupper-0.4) .. (gate_B.west);
	\draw[curvearrow, green] (gate_B.east) .. controls (5.5, \yupper-0.4) and (3.5, \yupper+0.4) .. (gate_A.west);
	\draw[curvearrow, red] (gate_C.east) .. controls (5.5, \ylower+0.4) and (3.5, \ylower-0.4) .. (gate_D.west);
	\draw[curvearrow, green] (gate_D.east) .. controls (5.5, \ylower-0.4) and (3.5, \ylower+0.4) .. (gate_C.west);

	\filldraw (main_control_source) circle (1.5pt);
	\filldraw (control_source_upper) circle (1.5pt);
	\draw (main_control_target) circle (2.5pt);
	\draw (control_target_upper) circle (2.5pt);
	\draw (control_target_lower) circle (2.5pt);
	
\end{tikzpicture}%
}%
}

%% file: diagram_soc.tex
\newcommand{\diagramsoc}{%
\resizebox{\columnwidth}{!}{%
  \begin{tikzpicture}[
	thick,
	gate/.style={draw, fill=gray!20, minimum height=0.6cm, minimum width=0.6cm},
	midarrow/.style={
		decoration={
			markings,
			mark=at position 0.5 with {\arrow{stealth}}
		},
		postaction={decorate}
	},
	curvearrow/.style={
		decoration={
			markings,
			mark=at position 0.7 with {\arrow{stealth}}
		},
		postaction={decorate}
	},
	revcurvearrow/.style={
		decoration={
			markings,
			mark=at position 0.7 with {\arrow{stealth[reversed]}}
		},
		postaction={decorate}
	}
]
	
	\node (rho_c) at (0, 3) {$\rho_c$};
	\node (rho_t) at (0, 0) {$\rho_t$};
	\coordinate (main_control_source) at (1, 3);
	\coordinate (main_control_target) at (1, 0);
	\coordinate (main_fork) at (1.5, 0);
	\coordinate (main_rejoin) at (7.5, 0);
	\node (rho_out) at (9, 0) {$\rho_{out}$};

	\pgfmathsetmacro{\yupper}{1.5}
	\coordinate (control_source_upper) at (2.5, 3) {};
	\coordinate (control_target_upper) at (2.5, \yupper) {};
	\coordinate (fork_upper) at (3, \yupper);
	\coordinate (rejoin_upper) at (6, \yupper);
	\node[gate] (gate_A) at (4.5, \yupper+0.75) {$\boldsymbol{\mathcal{E}^{(1)}}$};
	\node[gate] (gate_B) at (4.5, \yupper-0.75) {$\boldsymbol{\mathcal{E}^{(2)}}$};

	\pgfmathsetmacro{\ylower}{-1.5}
	\coordinate (control_target_lower) at (2.5, \ylower) {};
	\coordinate (fork_lower) at (3, \ylower);
	\coordinate (rejoin_lower) at (6, \ylower);
	\node[gate] (gate_C) at (4.5, \ylower+0.75) {$\boldsymbol{\mathcal{E}^{(3)}}$};
	\node[gate] (gate_D) at (4.5, \ylower-0.75) {$\boldsymbol{\mathcal{E}^{(4)}}$};

	\draw (rho_c.east) -- (main_control_source) -- (control_source_upper) -- (9, 3);
	\draw (main_control_source) -- (main_control_target);
	\draw (control_source_upper) -- (control_target_upper);
	\draw (control_source_upper) -- (control_target_lower);

	\draw[midarrow] (rho_t.east) -- (main_control_target);
	\draw[-stealth] (main_control_target) -- (main_fork);
	
	\draw[midarrow] (main_fork) |- (control_target_upper);
	\draw[midarrow] (main_fork) |- (control_target_lower);
	
	\draw[-stealth] (control_target_upper) -- (fork_upper);
	\draw[midarrow] (fork_upper) |- (gate_A.west);
	\draw[midarrow] (gate_A.east) -| (rejoin_upper);
	\draw[midarrow] (fork_upper) |- (gate_B.west);
	\draw[midarrow] (gate_B.east) -| (rejoin_upper);
	\draw[midarrow] (rejoin_upper) -| (main_rejoin);
	\draw[-stealth] (control_target_lower) -- (fork_lower);
	\draw[midarrow] (fork_lower) |- (gate_C.west);
	\draw[midarrow] (gate_C.east) -| (rejoin_lower);
	\draw[midarrow] (fork_lower) |- (gate_D.west);
	\draw[midarrow] (gate_D.east) -| (rejoin_lower);
	\draw[midarrow] (rejoin_lower) -| (main_rejoin);
	
	\draw[midarrow] (main_rejoin) -- (rho_out.west);

	\draw[revcurvearrow, blue] (fork_upper) .. controls (1.5, 0) and (7.5, 0) .. (rejoin_lower);
	\draw[curvearrow, yellow!80!black] (rejoin_upper) .. controls (7.5, 0) and (1.5, 0) .. (fork_lower);

	\filldraw (main_control_source) circle (1.5pt);
	\filldraw (control_source_upper) circle (1.5pt);
	\draw (main_control_target) circle (2.5pt);
	\draw (control_target_upper) circle (2.5pt);
	\draw (control_target_lower) circle (2.5pt);
	
\end{tikzpicture}%
}%
}

%% file: diagram_sos.tex
\newcommand{\diagramsos}{%
\resizebox{\columnwidth}{!}{%
  \begin{tikzpicture}[
	thick,
	gate/.style={draw, fill=gray!20, minimum height=0.6cm, minimum width=0.6cm},
	midarrow/.style={
		decoration={
			markings,
			mark=at position 0.5 with {\arrow{stealth}}
		},
		postaction={decorate}
	},
	curvearrow/.style={
		decoration={
			markings,
			mark=at position 0.7 with {\arrow{stealth}}
		},
		postaction={decorate}
	},
	revcurvearrow/.style={
		decoration={
			markings,
			mark=at position 0.7 with {\arrow{stealth[reversed]}}
		},
		postaction={decorate}
	}
]
	
	\node (rho_c) at (0, 3) {$\rho_c$};
	\node (rho_t) at (0, 0) {$\rho_t$};
	\coordinate (main_control_source) at (1, 3);
	\coordinate (main_control_target) at (1, 0);
	\coordinate (main_fork) at (1.5, 0);
	\coordinate (main_rejoin) at (7.5, 0);
	\node (rho_out) at (9, 0) {$\rho_{out}$};

	\pgfmathsetmacro{\yupper}{1.5}
	\coordinate (control_source_upper) at (2.5, 3) {};
	\coordinate (control_target_upper) at (2.5, \yupper) {};
	\coordinate (fork_upper) at (3, \yupper);
	\coordinate (rejoin_upper) at (6, \yupper);
	\node[gate] (gate_A) at (4.5, \yupper+0.75) {$\boldsymbol{\mathcal{E}^{(1)}}$};
	\node[gate] (gate_B) at (4.5, \yupper-0.75) {$\boldsymbol{\mathcal{E}^{(2)}}$};

	\pgfmathsetmacro{\ylower}{-1.5}
	\coordinate (control_target_lower) at (2.5, \ylower) {};
	\coordinate (fork_lower) at (3, \ylower);
	\coordinate (rejoin_lower) at (6, \ylower);
	\node[gate] (gate_C) at (4.5, \ylower+0.75) {$\boldsymbol{\mathcal{E}^{(3)}}$};
	\node[gate] (gate_D) at (4.5, \ylower-0.75) {$\boldsymbol{\mathcal{E}^{(4)}}$};

	\draw (rho_c.east) -- (main_control_source) -- (control_source_upper) -- (9, 3);
	\draw (main_control_source) -- (main_control_target);
	\draw (control_source_upper) -- (control_target_upper);
	\draw (control_source_upper) -- (control_target_lower);

	\draw[midarrow] (rho_t.east) -- (main_control_target);
	\draw[-stealth] (main_control_target) -- (main_fork);
	
	\draw[midarrow] (main_fork) |- (control_target_upper);
	\draw[midarrow] (main_fork) |- (control_target_lower);
	
	\draw[-stealth] (control_target_upper) -- (fork_upper);
	\draw[midarrow] (fork_upper) |- (gate_A.west);
	\draw[midarrow] (gate_A.east) -| (rejoin_upper);
	\draw[midarrow] (fork_upper) |- (gate_B.west);
	\draw[midarrow] (gate_B.east) -| (rejoin_upper);
	\draw[midarrow] (rejoin_upper) -| (main_rejoin);
	\draw[-stealth] (control_target_lower) -- (fork_lower);
	\draw[midarrow] (fork_lower) |- (gate_C.west);
	\draw[midarrow] (gate_C.east) -| (rejoin_lower);
	\draw[midarrow] (fork_lower) |- (gate_D.west);
	\draw[midarrow] (gate_D.east) -| (rejoin_lower);
	\draw[midarrow] (rejoin_lower) -| (main_rejoin);
	
	\draw[midarrow] (main_rejoin) -- (rho_out.west);

	\draw[revcurvearrow, blue] (fork_upper) .. controls (1.5, 0) and (7.5, 0) .. (rejoin_lower);
	\draw[curvearrow, yellow!80!black] (rejoin_upper) .. controls (7.5, 0) and (1.5, 0) .. (fork_lower);

	\draw[curvearrow, red] (gate_A.east) .. controls (5.7, \yupper+0.6) and (3.3, \yupper-0.6) .. (gate_B.west);
	\draw[curvearrow, green] (gate_B.east) .. controls (5.7, \yupper-0.6) and (3.3, \yupper+0.6) .. (gate_A.west);
	\draw[curvearrow, red] (gate_C.east) .. controls (5.7, \ylower+0.6) and (3.3, \ylower-0.6) .. (gate_D.west);
	\draw[curvearrow, green] (gate_D.east) .. controls (5.7, \ylower-0.6) and (3.3, \ylower+0.6) .. (gate_C.west);

	\filldraw (main_control_source) circle (1.5pt);
	\filldraw (control_source_upper) circle (1.5pt);
	\draw (main_control_target) circle (2.5pt);
	\draw (control_target_upper) circle (2.5pt);
	\draw (control_target_lower) circle (2.5pt);
	
\end{tikzpicture}%
}%
}

%% file: plotlegend.tex
\newcommand{\plotlegend}{%
  \begin{tikzpicture}[
    font=\sffamily\fontsize{5}{6}\selectfont, 
    every node/.style={anchor=west}
  ]
  \draw[green!50!black, thick] (0, 13.0) -- (0.5, 13.0);
  \node[green!50!black] at (0.55, 13.0) {Quantum Switch};
  \draw[violet, thick, dash pattern=on 5pt off 2pt] (0, 12.8) -- (0.5, 12.8);
  \node[violet] at (0.55, 12.8) {Coherent Superposition};
  \draw[black, thick] (0, 12.6) -- (0.5, 12.6);
  \node[black] at (0.55, 12.6) {SoS};
  \draw[red, thick, dash dot] (0, 12.4) -- (0.5, 12.4);
  \node[red] at (0.55, 12.4) {SoC};
  \draw[blue, thick, dash pattern=on 2pt off 4pt] (0, 12.2) -- (0.5, 12.2);
  \node[blue] at (0.55, 12.2) {CoS};
  \draw[orange, thick, dotted] (0, 12.0) -- (0.5, 12.0);
  \node[orange] at (0.55, 12.0) {CoC};
\end{tikzpicture}%
}